\documentclass[%
 reprint,
superscriptaddress,
 amsmath,amssymb,
 aps,
pra,
floatfix,
]{revtex4-2}
\usepackage[english]{babel}
\usepackage{graphicx}% Include figure files
\usepackage{dcolumn}% Align table columns on decimal point
\usepackage{bm}% bold math
\usepackage{mathdots}
\usepackage{color}
\usepackage{xcolor}
\usepackage{physics}
\usepackage{amsmath}
\usepackage{ragged2e}
\usepackage[normalem]{ulem}
\usepackage[T1]{fontenc}
\usepackage[utf8]{inputenc}
\usepackage{float}
\usepackage{stmaryrd}

\renewcommand{\vec}[1]{\mathbf{#1}}

\begin{document}

\preprint{APS/123-QED}

\title{{Non-classical optimization {of entangled photons} through complex media}}

\author{Baptiste Courme}%
\affiliation{% 
Sorbonne Université, CNRS, Institut des NanoSciences de Paris, INSP, F-75005 Paris, France}%
\affiliation{%
Laboratoire Kastler Brossel, ENS-Universite PSL, CNRS, Sorbonne Universite, College de France, 24 rue Lhomond, 75005 Paris, France 
}%

\author{Chloé Vernière}
 \affiliation{Sorbonne Université, CNRS, Institut des NanoSciences de Paris, INSP, F-75005 Paris, France}%Lines break automatically or can be forced with \\

\author{Malo Joly}%
\affiliation{%
Laboratoire Kastler Brossel, ENS-Universite PSL, CNRS, Sorbonne Universite, College de France, 24 rue Lhomond, 75005 Paris, France 
}%

\author{Daniele Faccio}
 \affiliation{School of Physics and Astronomy,University of Glasgow, Glasgow G12 8QQ, UK}%Lines break automatically or can be forced with \\

 \author{Sylvain Gigan}
 \affiliation{%
Laboratoire Kastler Brossel, ENS-Universite PSL, CNRS, Sorbonne Universite, College de France, 24 rue Lhomond, 75005 Paris, France 
}%
\author{Hugo Defienne}
 \affiliation{Sorbonne Université, CNRS, Institut des NanoSciences de Paris, INSP, F-75005 Paris, France}%Lines break automatically or can be forced with \\

\date{\today}

\begin{abstract}

Optimization approaches are ubiquitous in physics. In optics, they are key to manipulating light through complex media, enabling applications ranging from imaging to photonic simulators. In most demonstrations, however, the optimization process is implemented using classical coherent light, leading to a purely classical solution. Here we introduce the concept of optical non-classical optimization in complex media. We experimentally demonstrate the control and refocusing of non-classical light - namely, entangled photon pairs - through a scattering medium by directly optimizing the output coincidence rate. Importantly, these optimal solutions cannot be obtained with classical light, and do not result in a focus for classical light, a result of entanglement in the input state. {This genuinely non-classical optimization method promises advances in quantum imaging, communication, and optical simulations with non-classical light.}
%Beyond quantum imaging and communication, this genuinely non-classical optimization method shows strong potential for optical simulations with non-classical light.
\end{abstract}
\maketitle

\section{\label{sec:level1}Introduction}

\begin{figure*} [ht!]
    \centering
    \includegraphics[width = 1\textwidth]{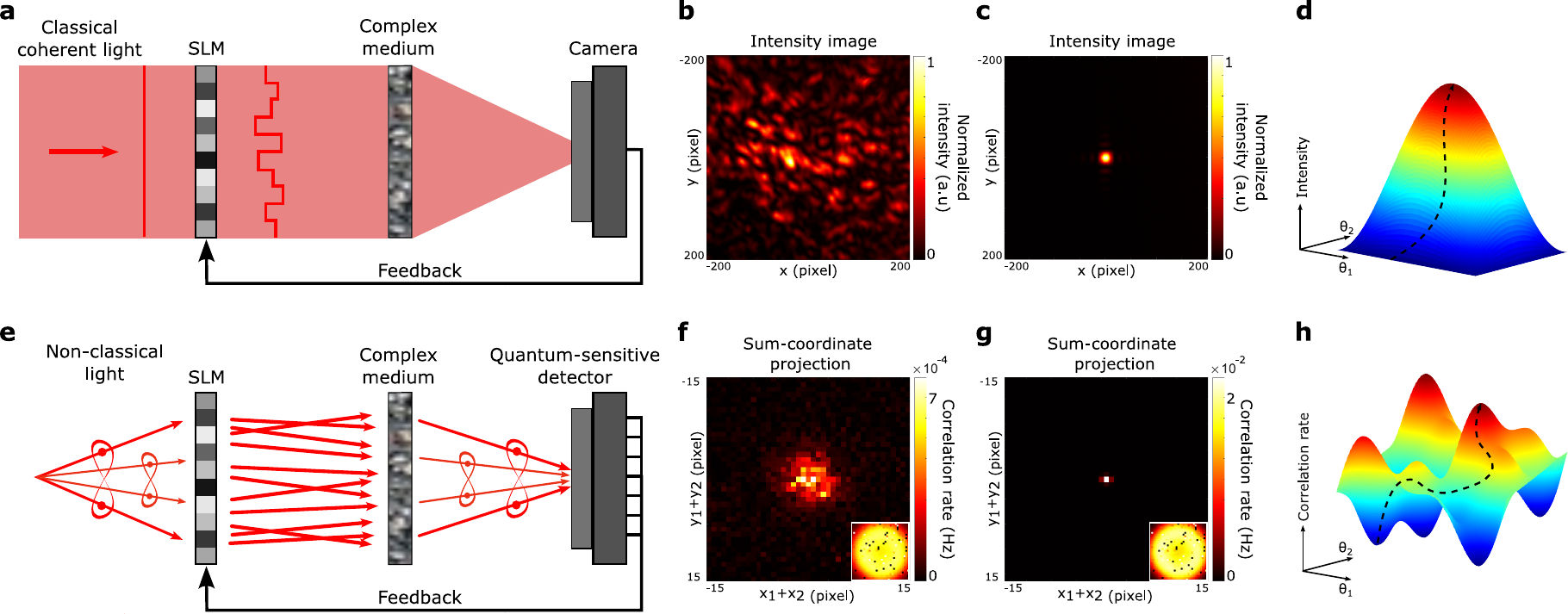}
    \caption{\textbf{Classical and non-classical optimizations.} \textbf{a,} Refocusing classical coherent light through a complex medium is achieved by optimization-based wavefront shaping, as described in Ref.~\cite{vellekoop_focusing_2007}. A feedback loop is established between the intensity of the output speckle pattern, measured by a camera, and a spatial light modulator (SLM) that shapes the wavefront of the incident beam. This iterative optimization increases intensity at the target spatial mode, effectively correcting distortions introduced by the medium. \textbf{b and c,} Intensity images measured before and after after optimization, respectively. \textbf{d,} Optimization landscape showing a unique maximum, where $\theta_1$ and $\theta_2$ represent the phases of two SLM pixels. \textbf{e,} In non-classical optimization, a feedback loop is established between a non-classical property of the quantum state measured at the output and a shaping device at the input. This loop iteratively adjusts the input shaping to enhance the targeted non-classical property. In our experiment, non-classical optimization is implemented using spatially-entangled photon pairs propagating through a scattering medium and detected in coincidence using a single-photon avalanche diode (SPAD) camera. Photons are strongly correlated in their momentum i.e. between \(\vec{k}\) and \(-\vec{k}\). \textbf{f,} Before optimization, the sum-coordinate projection of $\Gamma$, noted $\Gamma^+$, exhibits a speckle pattern, while the intensity is uniform (inset). \textbf{g,} If the optimization is successful, a peak emerges in $\Gamma^+$, confirming the restoration of strong spatial correlations between the photons, while the intensity remains uniform (inset). \textbf{h,} Optimization landscape showing multiple local maxima, where \(\theta_1\) and \(\theta_2\) represent the phases of two SLM pixels.}
    \label{fig 1}
\end{figure*}

In 2007, Vellekoop and Mosk demonstrated the focusing of classical coherent light through opaque scattering media~\cite{vellekoop_focusing_2007}. They developed an optimization process that taillors the wavefront of a coherent beam incident on the medium using a spatial light modulator (SLM), based on a speckle intensity feedback measured by a camera behind it (Figs.~\ref{fig 1}a-d). Two decades later, this breakthrough has led to numerous applications in deep-tissue imaging~\cite{horstmeyer_guidestar_2015,gigan_roadmap_2022}, light transport in multimode fibers~\cite{vcivzmar_shaping_2011}, optical trappings~\cite{taylor_enhanced_2015} and complex problems simulation~\cite{pierangeli_scalable_2021,leonetti_optical_2021}. In recent years, it was also extended to quantum optics ~\cite{lib_quantum_2022}. {Pioneering studies have, for example, reported the focusing of single photons and photon pairs through {thick scattering media}~\cite{huisman_controlling_2014,defienne_nonclassical_2014,wolterink_programmable_2016} and {multimode fibers}~\cite{carpenter_110x110_2014,defienne_two-photon_2016}. These approaches have subsequently been applied to more practical scenarios, such as programming quantum circuits~\cite{leedumrongwatthanakun_programmable_2020,goel_inverse_2024,makowski_large_2024}, imaging though optical aberrations~\cite{cameron_quantum-assisted_2023}, and restoring entanglement after transmitting photons through fibers~\cite{valencia_unscrambling_2020}, scattering layers~\cite{lib_real-time_2020,devaux_restoring_2023,courme_manipulation_2023,shekel_shaping_2024} and turbulent atmosphere~\cite{gruneisen_adaptive_2021,zhao_performance_2020,scarfe_fast_2023}. }

Although all these studies involve non-classical optical states, they still rely on classical wavefront shaping. Indeed, in most of them, wavefront optimization {is performed first} using a classical light source - a laser aligned in the same direction as the quantum state, sharing similar spectral and polarization properties - {before applying the optimal solution to the quantum signal}. This method benefits from an intense signal for optimization, which is particularly useful in high-loss scenarios (e.g. multiple scattering) and dynamic conditions (e.g. turbulence)~\cite{vellekoop_focusing_2007,popoff_measuring_2010,stockbridge_focusing_2012}. However, it completely ignores the non-classical features of the quantum state involved, making the process purely classical.

Ignoring the quantum state's specifics in the shaping process has its limits. Practically, ensuring perfect mode matching between a classical and a non-classical source is often challenging. For example, in the context of high-dimensional entangled states, such as spatially entangled photon pairs~\cite{walborn_spatial_2010}, scattering has only been corrected either simultaneously for a small subset of modes~\cite{defienne_adaptive_2018,devaux_restoring_2023,courme_manipulation_2023} or sequentially by addressing each mode individually~\cite{valencia_unscrambling_2020,goel_inverse_2024}. More importantly, it fundamentally restricts the efficiency and range of the control achieved. Purely non-classical phenomena, such as complex quantum interference~\cite{ott_quantum_2010,makowski_large_2024} and entanglement scrambling~\cite{valencia_unscrambling_2020,courme_manipulation_2023}, for example, are not taken into account when using classical optimization. 
%Optimizing with direct feedback from a measurement of the non-classical state - \textit{non-classical optimization} - could yield new solutions specific to the quantum state, unattainable through classical optimization.}

Here, we present an optimization-based wavefront shaping method with direct feedback from a non-classical feature of a quantum state, leading to solutions unattainable through classical optimization - {a method we refer to as `non-classical optimization'}. Our experimental implementation involves an entangled two-photon state propagating through a scattering medium. Contrary to previous works, the feedback signal is derived directly from photon coincidence measurements across spatial modes at the output, which are specific to the non-classical source. 
To better understand this behavior, we investigate the two-photon interference processes occurring in the scattering medium and {develop} an analytical model for the the spatial correlation modulation. 
We demonstrate that the optimization converges towards unique solutions which do not refocus classical light. Crucially, we show that these non-classical solutions exist and are accessible only if entanglement is present between the photons. {Finally, we discuss the potential of this non-classical optimization approach beyond imaging and communication, such as in solving complex spin-glass optimization problems.}

\section{Concept}

{Figure~\ref{fig 1} conceptually compares classical and non-classical optimization. A non-classical state propagates through a complex medium and is measured at the output by a detector sensitive to its quantum properties, such as a single-photon or homodyne detector. A feedback mechanism is then implemented, linking this measurement to a programmable device that shapes the properties of the state incident on the medium. An iterative algorithms adjust the input (usually the spatial phase pattern) towards a configuration that optimizes the targeted non-classical property.

% In our work, non-classical optimization is demonstrated using a spatially-entangled two-photon state propagating through a scattering layer and detected by a single-photon avalanche diode (SPAD) camera. 
In our work, non-classical optimization is demonstrated using a spatially-entangled two-photon state propagating through a scattering layer and detected by a single-photon avalanche diode (SPAD) camera.  At the input, the state has strong multimode spatial correlations: when a photon is detected in any transverse spatial mode $\vec{k}$, its twin has a high probability of being detected precisely in the symmetric mode $-\vec{k}$~\cite{walborn_spatial_2010}. At the output, the SPAD camera detects photons across all spatial modes in parallel, enabling measurement of both the intensity image (single counts) and the spatially-resolved second-order correlation function $\Gamma$ (coincidence counts)~\cite{ndagano_imaging_2020}. After propagation, the intensity remains uniform, but a speckle pattern appears in the sum-coordinate projection {of} $\Gamma$ , {noted $\Gamma^+$} (Fig.~\ref{fig 1}f). This speckle results from complex two-photon interference~\cite{peeters_observation_2010} and indicates a degradation of the initial strong spatial correlation between the photons. The correlation value between any arbitrary pair of pixels is chosen as the optimization target, and a feedback loop is implemented with a SLM that modulates the wavefront of the input state. If the optimization is successful, a peak emerges in $\Gamma^+$ (Fig.~\ref{fig 1}g).

\section{{Optimization}}

% The situation described in Figure~\ref{fig 1} falls into a more general optimization problem.
{When a pure two-photon state propagates through a linear scattering medium, the coincidence rate $\Gamma_{kl}$ between two output modes $k$ and $l$ can be expressed as 
\begin{equation}
    \Gamma_{kl} = \left | \sum_{m,n} t_{km} t_{ln} \psi_{mn} e^{i \theta_{nm}} \right |^2,
    \label{generalformula}
\end{equation}
where $\psi_{mn}$ is the input two-photon field decomposed over the input modes $m$ and $n$, $t_{kn}$ is a scattering matrix coefficient linking input mode $n$ to output mode $k$, and \( \theta_{nm} \) are adjustable phase terms. In our study, we consider only spatial modes (photons have the same frequency and polarization), though the problem can generalize to all degrees of freedom. It is clear that a maximum value of $\Gamma_{kl}$ is reached when all the terms in Equation~\eqref{generalformula} are in phase i.e. when all $\theta_{nm}$ are carefully adjusted to cancel the phase of $t_{km} t_{ln} \psi_{mn}$. This problem is analogous to the classical optimization problem detailed in Ref.~\cite{vellekoop_focusing_2007}, which is well known to have a unique solution (up to a global phase) with only a global maximum and no local minima, as shown in Figure~\ref{fig 1}d and demonstrated in Supplementary Section I. See also Methods for details on Equation~\eqref{generalformula}.

For coincidence optimization, unlike the classical case, a single SLM plane does not have enough degrees of freedom to independently control all the \(\theta_{nm}\) and perfectly align all the terms in Equation~\eqref{generalformula}. For instance, considering only two spatial input modes (SLM pixels \(m\) and \(n\)) and two spatial output modes (camera pixels \(k\) and \(l\)), three terms corresponding to four two-photon paths must be put in phase: \(t_{km} t_{lm} \psi_{mm} e^{2 i \theta_m}\) (Fig.~\ref{fig 2}a), \(t_{ln} t_{kn} \psi_{nn} e^{2 i \theta_n}\) (Fig.~\ref{fig 2}b) and \( (t_{kn} t_{lm} + t_{km} t_{ln}) \psi_{mn} e^{i (\theta_m + \theta_n)}\) (Figs.~\ref{fig 2}c and d), where $\theta_m$ and $\theta_n$ are phase shifts applied to modes $n$ and $m$.  When considering \(N > 2\) input modes, the optimization problem thus becomes finding a set of phases \( \{ \theta_n \} \), \(n \in [1, N]\), that optimally adjusts the \(N^2\) phases \(\theta_{nm}\) to maximize the coincidence rate between the target output modes. This optimization problem is significantly more complex than its classical counterpart. For instance, simple simulations shown in Figure~\ref{fig 1}h and in Supplementary Section II demonstrate the presence of local maxima and clearly indicate that the optimal solution is not always unique.} 

To tackle this problem, we have adapted an iterative optimization approach derived from the random partition algorithm~\cite{vellekoop_phase_2008}. At each step, a subset of SLM pixels are selected and shifted (active SLM area) relative to another part that remains flat (reference SLM area). The two-photon speckle fields interfere at the output, causing the target coincidence rate $\Gamma_{kl}$ to vary as
\begin{equation}
    \Gamma_{kl}(\theta) = A \, cos (2 \theta + \theta_A)+B \, cos(\theta+\theta_B) + C,
    \label{generalInter}
\end{equation}
{where \( A \), \( B \), \( C \), \( \theta_A \), and \( \theta_B \) are amplitude and phase coefficients that depend on the input two-photon state and scattering matrix, $\theta$ is the relative global phase between the active and reference SLM areas, $k$ and $l$ are two arbitrarily chosen output spatial modes with transverse positions $(x_k,y_k)$ and $(x_l,y_l)$. In Equation~\eqref{generalInter}, the \( 2\theta \) oscillation arises from the two-photon paths linking one input to two outputs (Figs.~\ref{fig 2}a and b), while the \( \theta \) oscillation originates from paths linking two inputs to two outputs (Figs.~\ref{fig 2}c-d). Figure~\ref{fig 2}e shows an example of $\Gamma_{kl}$ function of $\theta$ measured experimentally using the setup described later in Figure~\ref{fig 2}f.
The value of $\theta$ maximizing \(\Gamma_{kl}\) is determined by phase-shifting digital holography or by fitting the data with Equation~\eqref{generalInter}, and is then applied to the pixels in the active area. The algorithm proceeds by selecting and shifting a new subset of pixels at each step, repeating this process until the optimization target no longer evolves. The demonstration of Equation~\eqref{generalInter} and the search for its optimal phase are detailed in Methods and Supplementary Section III.

\begin{figure*} [ht!]
    \centering
    \includegraphics[width = 1
\textwidth]{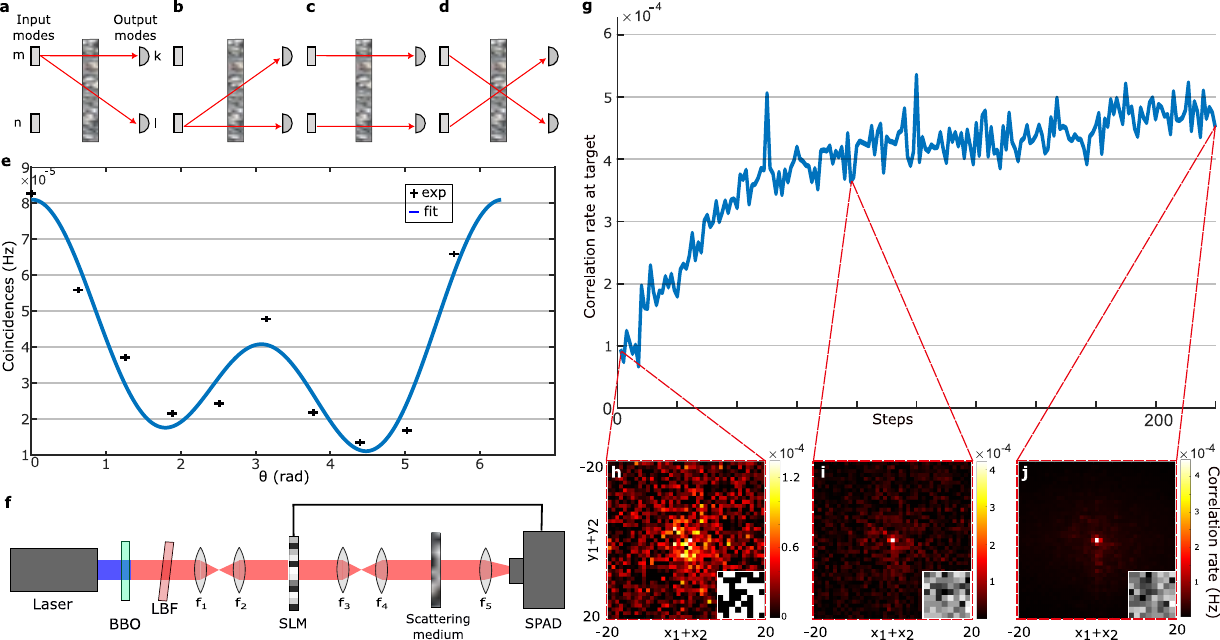}
   \caption{\textbf{Optimization results.} {Four two-photon interference paths occur between two input modes (SLM pixels) and two output modes (SPAD pixels): \textbf{a and b,} Two photons originating from a single input mode $m$ or $n$ and detected at each ouptut modes $k$ and $l$. \textbf{c,} Two photons originating from two input modes $m$ and $n$ detected at output modes $k$ and $l$, and \textbf{(d)} vice versa. 
   \textbf{e,} Experimental measurement of $\Gamma_{kl}$ between two arbitrary output modes $k$ and $l$ as a function of the phase \( \theta \) between the active and reference areas of the SLM. $10$ points are measured (black crosses) and fitting the data (blue curve) enables to determine the coefficients A=$2.21\cdot10^{-5}$Hz, B=$2.0\cdot10^{-5}$Hz, C=$3.87\cdot10^{-5}$Hz, $\theta_A$=0.03 rad, $\theta_B$=-0.14 rad $R^2 = 0.93 $) and an optimal phase value of approximately $0.03$ rad.} \textbf{f,} {A collimated diode laser ($405$ nm) illuminates $\beta$-barium borate (BBO) crystal with a thickness of $0.5$mm to produce spatially entangled pairs of photons by type I spontaneous parametric down conversion (SPDC). A long-pass (LPF) positioned after the crystal remove pump photons. Lenses $f_1-f_2$ image the crystal surface onto a spatial light modulator (SLM), that is itself imaged by lenses $f_3-f_4$ onto an optical plane where a scattering medium is positioned. The distance between two successive lenses equals the sum of their focal lengths. At the output, a lens $f_5$ and a single-photon avalanche diode (SPAD) camera positioned in its focal plane collect the photons. A band-pass filter $810 \pm 5$nm is positioned in front of the camera (not represented).} \textbf{g,} {Optimization curve showing the evolution of the correlation value at the target point as a function of the number of steps taken in the optimization procedure.} The enhancement factor is {approximately} $5$. {The optimization was performed over $200$ steps, corresponding to approximately $120$ hours.} \textbf{h-j,} $\Gamma^+$ and SLM phase patterns (insets) at three different steps of the process. More details on the experimental setup can be found in Supplementary Section IV.}
    \label{fig 2}
\end{figure*}

\section{Experiment}

This algorithm is implemented using the experimental setup described in Figure~\ref{fig 2}f. Pairs of photons at \( 810 \) nm with the same polarization are generated via type-I spontaneous parametric down-conversion (SPDC) in a thin \(\beta\)-barium borate (BBO) crystal illuminated by a collimated continuous-wave \( 405 \) nm laser. The resulting two-photon state is entangled in high spatial dimensions~\cite{walborn_spatial_2010}. The crystal output surface is imaged onto an SLM using a $4f$ telescope, {which is itself imaged onto an optical plane near where a thing scattering medium (layer of parafilm) is placed.}
{In the SLM plane (input plane), the two-photon wavefunction $\psi$ is characterized experimentally and modeled as a double Gaussian distribution~\cite{fedorov_gaussian_2009,schneeloch_introduction_2016}, with position and momentum correlation widths of $\sigma_{\vec{r}} = 2.9 \times 10^{-5} \, \mathrm{m}$ and $\sigma_{\vec{k}} = 8.0 \times 10^{2} \, \mathrm{m}^{-1}$, respectively.}
At the output, the photons are collected by a lens and a SPAD camera. The optical elements are arranged so that, in the absence of a medium, a Fourier plane of the crystal is imaged onto the SPAD. {In this configuration, photons are thus detected in the momentum basis, where they exhibit strong anti-correlations. This is evidenced by the sharp correlation peak in the sum-coordinate projection measured without the medium, shown in Figure~\ref{fig 3}a.} Further details on the experimental setup and input state characterization are provided in the Methods and Supplementary Sections IV and V.

{Figure~\ref{fig 2}g shows an optimization curve} obtained by modulating the incident wavefront while monitoring a coincidence target over $200$ steps. {At the input, the SLM is divided into $8 \times 8$ macro-pixels, each {37}$\times${37} SLM pixels. At the output, $\Gamma$ is measured and projected along the sum-coordinate axis to obtain $ \Gamma^+$ (see Methods). The optimization target is the coincidence value at an arbitrary pixel \( T \) of \( \Gamma^+ \). 
From the SPAD camera perspective, the value $\Gamma^{+}_T$ corresponds to the sum of coincidence rates between all symmetric pixel pairs around a pixel at position \((x_T, y_T)\) i.e. \( \Gamma^+_T = \sum_{k=1}^M \Gamma_{T+k,T-k} \), where \( T-k \) (resp. \( T+k \)) corresponds to a pixel at \((x_T-x_k, y_T-y_k)\) (resp. \((x_T+x_k, y_T+y_k)\)), and \( M \) is the total number of pixels. {In our experimental configuration, maximizing \(\Gamma^+_T\) instead of \(\Gamma_{kl}\) for a given pair of spatial modes $k$ and $l$ has the practical advantage of yielding to stronger signal-to-noise ratio (SNR), while preserving the modulation shape established in Equation~\eqref{generalInter}, as shown in Methods and Supplementary Section VI. Simulations shown in Supplementary Section IX of the manuscript also present non-classical optimizations performed with $\Gamma_{kl}$ as the target.} }

At each optimization step, half of the macro-pixels are randomly chosen and shifted using $6$ or $9$ phase increments between $0$ and $2 \pi$, respectively for holography or fitting (see Methods and Supplementary Section III). Each data point in Figure~\ref{fig 2}e corresponds to a 5-minute acquisition. Values of \( \Gamma^+_T \) for different phases $\theta$ are then fitted using the model in Equation~\eqref{generalInter} to determine the optimal phase shift, which is then applied to the SLM active area.  Figures~\ref{fig 2}h-j show the images \( \Gamma^+ \) at various steps, illustrating the emergence of a peak at the target position, which confirms successful refocusing of photon-pair correlations after the scattering medium. 
{Note that the presented results were obtained by controlling only $8 \times 8 = 64$ input modes, offering a good experimental trade-off between enhancement and acquisition time. As detailed in Ref.~\cite{shekel2025fundamentalboundswavefrontshaping} and confirmed experimentally in Supplementary Section~VII, increasing the number of modes leads to a linear improvement in the enhancement ratio of the final correlation image. However, it also increases the number of steps required for convergence, and thus the total acquisition time. This limitation could be easily mitigated in future implementations by using faster and more efficient coincidence counting cameras, such as the Tpxcam sensor~\cite{nomerotski_imaging_2019,courme_quantifying_2023}, as shown in Supplementary Section~VIII.}
 
\begin{figure*} [ht!]
    \centering
    \includegraphics[width = 1\textwidth]{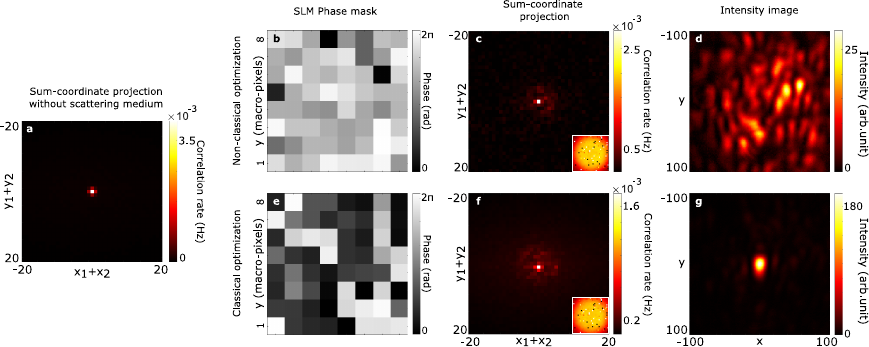}
\caption{\textbf{Focusing results.} {Sum-coordinate projection acquired without the scattering medium \textbf{(a)}}. SLM phase mask \textbf{(b)} and sum-coordinate projection \textbf{(c)} obtained after refocusing spatial correlations {through the scattering medium} using non-classical optimization with a two-photon entangled state at the input. The output intensity image is shown in inset. Speckle intensity \textbf{(f)} observed at the output after replacing the two-photon state with classical coherent light. SLM phase mask \textbf{(e)} obtained through classical optimization. In this case, classical light is first refocused in the output intensity \textbf{(g)}. After replacing the classical source with entangled photon pairs, refocusing is then also observed in the sum-coordinate projection \textbf{(f)}. The output intensity image is shown in the inset.}.
    \label{fig 3}
\end{figure*}

Ultimately, the optimization converges towards the phase mask shown in Figure~\ref{fig 3}b. The focusing peak shown in Figure~\ref{fig 3}c is approximately $5$ times more intense than the average two-photon speckle surrounding it, while the direct intensity image remains homogeneous (inset). Interestingly, when replacing the photon pair source with a laser of the same polarization and frequency, we do not observe any refocusing of the classical light in the output intensity image in Figure~\ref{fig 3}d. Instead, a speckle pattern appears. This phase mask thus mitigates scattering for entangled photons, but not for classical light.

This result is surprising, in particular when it is compared to the outcome obtained using a classical optimization approach. In this case, the intensity at the center of the output CCD camera is first maximized with the laser at the input using the method of Ref.~\cite{vellekoop_focusing_2007}, resulting in an intense peak visible in Figure~\ref{fig 3}g. Then, the classical source is replaced by photon pairs while maintaining the same phase mask on the SLM (Fig.~\ref{fig 3}e). A peak appears in the sum-coordinate projection in Figure~\ref{fig 3}f, with a value similar to that obtained using the non-classical optimization approach. Non-classical optimization thus converges to phase masks distinct from the classical solution, yet still refocuses entangled photons at the medium output. Additional simulations shown in Supplementary Section IX and X confirm that these results are reproducible in a range of parameters, including for thicker scattering media (e.g. multiple layers of parafilm stacked together). 

{{\section{Role of entanglement}}

The existence of these `non-classical' solutions i.e. phase masks that refocus entangled photons without refocusing classical light, is intrinsically linked to the presence of entanglement in the input state. To support this claim, we compare our optimization results with those obtained using non-entangled states. To ensure a meaningful comparison, these separable states are chosen to produce, in the absence of the scattering medium, similar spatial correlations in the camera plane as those observed experimentally with the entangled state. They are described by a density matrix $\rho =  \sum_{j} p_j \left | \psi^j \right \rangle  \left \langle \psi^j \right |$, where $ \ket{\psi^j}$ are separable two-photon states of factorizable wavefunction $\psi_{nm}^j = \phi^j_n \chi^j_m$, and $p_j$ its probability. After propagation through the medium, one can show that Equation~\eqref{generalformula} simplifies into
\begin{equation}
\Gamma_{kl} = \sum_{j} p_j \left | \sum_{n} t_{kn} \phi^j_{n} e^{i \theta_{n}} \right |^2 \left |  \sum_{m} t_{lm} \chi^j_{m} e^{i \theta_{m}} \right |^2.
\end{equation}
In this case, maximizing the correlations between output modes $k$ and $l$ is equivalent to maximizing the sum of output intensities products measured at modes $k$ and $l$ when an ensemble of coherent classical fields, $\phi^j$ and $\chi^j$, are propagated through the scattering medium.  
As detailed in Methods and Supplementary Section XI, we construct two examples of such separable states that reproduce the same sum-coordinate projection as this of the entangled state shown in Figure~\ref{fig 3}a.
As detailed in Supplementary Section XIII and Figure~S13, classical-light experiments and simulations show that non-classical optimization can be implemented with these separable states, restoring a correlation peak at the output.
Unlike the entangled case, however, a focus also appears at the center of the output classical intensity image, indicating that the optimization consistently converges to the classical solution. Thus, with a non-entangled state, non-classical optimization is equivalent to classical optimization.
 
However, it is important to note that while entanglement is a necessary condition for reaching a non-classical solution, it is not sufficient. Figure~\ref{fig_4_V2} presents simulation results of non-classical optimizations performed with entangled two-photon input states of varying degrees of entanglement.
These simulations use the same optical system as described in Figure~\ref{fig 2}f. The degree of entanglement, quantified by the Schmidt number $K$, is controlled by varying the spatial correlation width $\sigma_\vec{r}$ in the SLM plane, while keeping $\sigma_\vec{k}$ fixed (see Methods).
In this configuration, all input states produce identical sum-coordinate projections in the absence of scattering, allowing for a consistent basis for comparison. Figure~\ref{fig_4_V2} shows the similarity between each classical intensity pattern obtained after non-classical optimization and the one resulting from a direct classical optimization i.e. a similarity value close to one indicates convergence toward the same solution in both cases.
Interestingly, we observe that for values of $\sigma_\vec{r} > 240 \mu$m, corresponding to entangled states with $1 < K < 3$, the non-classical optimization converges primarily to the classical solution. This result shows that entanglement alone does not determine the emergence of non-classical solutions.
\begin{figure} [H]
    \centering
    \includegraphics[width = 1\columnwidth]{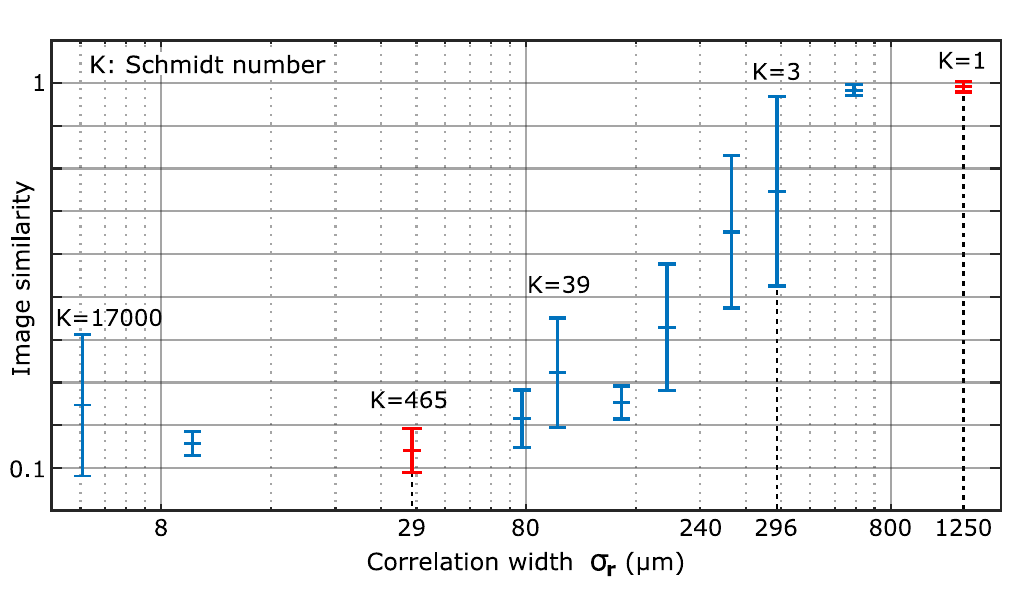}
\caption{{\textbf{Non-classical optimizations using entangled states with different degree of entanglement.} Simulation results of non-classical optimization using two-photon states with varying degrees of entanglement. The x-axis indicates the spatial correlation width $\sigma_\vec{r}$ in the SLM plane, and the y-axis shows the image similarity (computed using the \texttt{corr2} function in MATLAB) between the classical intensity pattern obtained after each non-classical optimization and that of the classical solution. A similarity value close to $1$ indicates convergence toward the classical solution, while a significantly lower value indicates a non-classical outcome. For each value of $\sigma_\vec{r}$, the optimization is repeated five times, and the error bars represent the standard deviation of the resulting similarity values. The two red data points correspond to the experimental cases shown in Figures~\ref{fig 3}b–d (high-dimensional entangled state, $K = 465$) and Supplementary Figures~S13a–d (pure separable state, $K = 1$). Selected Schmidt numbers $K$ are indicated above the corresponding points. A transition between convergence to classical and non-classical solutions appears when $\sigma_{\vec{r}}$ exceeds approximately $296,\mu$m, which corresponds to the size of one controlled mode on the SLM (i.e. a macropixel). See Methods for additional details on the simulations.}}
    \label{fig_4_V2}
\end{figure}

In our experiment, the convergence - or not - toward non-classical solutions, particularly the transition observed around $\sigma_\vec{r} = 240 \mu$m in Figure~\ref{fig_4_V2}, can be understood as a competition between first- and second-order coherence of the SPDC light at the input. As discussed in Ref.~\cite{saleh_duality_2000}, when the degree of entanglement increases (for a fixed SPDC beam width, set here by $\sigma_\vec{k}$), the first-order spatial coherence of the beam decreases.
In the high-Schmidt-number regime, first-order coherence is negligible, and the SLM primarily modulates second-order coherence, thereby acting on two-photon interference - the key mechanism responsible for producing non-classical solutions.
However, as the Schmidt number decreases, the first-order spatial coherence becomes sufficient for the SLM to influence single-photon interference, steering the optimization toward classical solutions.
In Figure~\ref{fig_4_V2}, we observe that the transition occurs precisely when $\sigma_{\vec{r}}$ becomes smaller than the SLM macropixel size (i.e. the size of an input mode), approximately $296\,\mu$m, which is thus consistent with our interpretation. The ratio between the SLM mode size and the spatial correlation width $\sigma_\vec{r}$ at the input plane is thus a critical parameter governing whether the optimization converges to a classical or non-classical solution.

\section{Discussion}

In this work, we implement a non-classical optimization approach using an entangled two-photon state propagating through a scattering medium, using as feedback for the SLM pattern the photon coincidence signal from a SPAD camera. We demonstrate the existence of specific `non-classical' solutions that enable the refocusing of entangled photon pairs without refocusing classical light. Remarkably, these solutions cannot be reached using classical optimization algorithms and have never been observed in previous studies.  
 
We show that the existence of these non-classical solutions is directly linked to the presence of entanglement. {Intuitively, this link stems from the ability of entangled states to maintain correlations across multiple measurement bases~\cite{spengler_entanglement_2012}.}
 During {non-classical optimization}, the SLM and scattering medium - forming a linear optical system - reconfigure to implement a spatial basis transformation that maximizes correlations. Classically, only a basis that approximates the identity transformation allows refocusing at the output of the medium, as the system effectively tries to make the scattering medium transparent. For entangled states, however, multiple bases become accessible, allowing the system to converge toward one of them, which does not necessarily correspond to the identity. {This multiplicity of bases can be interpreted as distinct maxima within a complex optimization landscape. The optimization process may converge to a local maximum when its value is comparable to that of the maximum associated with the classical solution. However, this scenario arises only when the degree of entanglement is sufficiently high, ensuring the existence of multiple alternative bases beyond the identity i.e. multiple non-classical maxima. In our experimental configuration, we show that the key parameter governing convergence toward non-classical solutions is the ratio between the size of a controlled SLM input mode and the correlation width $\sigma_{\vec{r}}$ in the SLM plane.}
 
{In terms of applications}, non-classical optimization is primarily a tool that helps mitigate external disturbances to a quantum state, a major challenge in quantum technologies. In optics, this issue arises in various practical scenarios, such as quantum imaging systems affected by optical aberrations~\cite{cameron_quantum-assisted_2023,freitas_quantum_2024}, quantum communication through turbulent atmospheres and fibers~\cite{acosta_analysis_2024,valencia_unscrambling_2020,gruneisen_adaptive_2021}, and even quantum computation algorithms where error correction is needed~\cite{muller_iterative_2016}. 
{Beyond that, our approach may also serve as a powerful tool for advancing optical simulation. Recent studies have demonstrated that classical optical optimization techniques can be used to find the ground states of complex physical systems, such as those described by Ising model Hamiltonians~\cite{pierangeli_large_2019,pierangeli_scalable_2021,PhysRevLett.134.063802}. Solving such problems is known to be computationally hard at large scale. However, these optical simulators have so far been limited to relatively simple Hamiltonians involving only spin-spin interactions.
In contrast, as we show in Methods and Supplementary Section~XII, our system can simulate more complex models described by Hamiltonians that include higher-order multi-spin (more than two) interactions. In this context, non-classical optimization enables the minimization of the system energy to identify its ground state. For example, this is demonstrated experimentally in Supplementary Figure~S12, where we simulate a $64$-spin system governed by a Hamiltonian containing two-spin, three-spin, and four-spin interactions, a level of complexity not accessible to classical optical Ising machines.
With further developments in programmability and improved convergence speed, our non-classical optimization framework can also become a new tool for solving new classes of computationally hard problems.
} }

\section*{Methods}

\noindent \textbf{Details on Equation~\eqref{generalformula}.} Considering photons degenerate in frequency and polarization, a pure two-photon state is expressed as  
\begin{equation}  
| \Psi\rangle = \iint \psi(\vec{r_1},\vec{r_2}) a^\dagger_{\vec{r_1}} a^\dagger_{\vec{r_2}} d\vec{r_1} d\vec{r_2} | 0 \rangle,  
\end{equation}  
where \(a^\dagger_{\vec{r_1}}\) is the photon creation operator in the spatial mode \(\vec{r_1}\), and \(\psi(\vec{r_1},\vec{r_2})\) represents the spatial two-photon wavefunction. {After propagation through a linear optical system characterized by a coherent point spread function $t(\vec{r'}, \vec{r})$, where $\vec{r}$ and $\vec{r'}$ denote transverse positions in the input and output planes, respectively, the output two-photon wavefunction, noted $\psi'$, becomes~\cite{abouraddy_entangled-photon_2002}:
\begin{equation}
\label{generalcontinue}
\psi'(\vec{r_1'},\vec{r_2'}) = \iint \psi(\vec{r_1},\vec{r_2}) t(\vec{r_1'},\vec{r_1})  t(\vec{r_2'},\vec{r_2}) d\vec{r_1} d\vec{r_2}.  
\end{equation}  
}
The spatial second-order correlation function $\Gamma$ is defined as the squared absolute value of the corresponding two-photon wavefunction i.e. $\Gamma(\vec{r_1}, \vec{r_2}) = |\psi(\vec{r_1}, \vec{r_2})|^2$. Equation~\eqref{generalformula} is thus derived by discretizing Equation~\eqref{generalcontinue}. Specifically, in the discretized representation, the function $t(\vec{r'}, \vec{r})$ corresponds to the scattering matrix $t_{kl}$, where $l$ labels an input pixel on the spatial light modulator (SLM) and $k$ labels an output pixel on the camera. \\
\\
\noindent \textbf{Details on Equation~\eqref{generalInter}.} Considering two output modes \( k \) and \( l \) and two SLM areas, denoted \( \mathcal{M} \) and \( \mathcal{N} \) (active and reference areas), the coincidence rate \( \Gamma_{kl} \) results from the coherent sum of complex terms associated with the two-photon interference processes shown in Figure~\ref{fig 2}a–d.
\begin{equation}
 \Gamma_{kl} = \left| e^{2i\theta_{\mathcal{M}}} S_{\mathcal{M}} + e^{2i\theta_{\mathcal{N}}} S_{\mathcal{N}} + e^{i(\theta_{\mathcal{M}}+\theta_{\mathcal{N}})} S_{\mathcal{M},\mathcal{N}} \right|^2,
 \label{equation2detail}
\end{equation}
where $\theta_\mathcal{M}$ and $\theta_\mathcal{N}$ are the phases of the active and reference areas and  
\begin{eqnarray}
    S_{\mathcal{M}} &=& \sum_{m\in \mathcal{M}} t_{km} t_{lm} \psi_{mm}, \\
    S_{\mathcal{N}} &=& \sum_{n\in \mathcal{N}} t_{kn} t_{ln} \psi_{nn},  \\
    S_{\mathcal{M},\mathcal{N}} &=& \sum_{m\in \mathcal{M},n\in \mathcal{N}} [t_{kn} t_{lm} + t_{km} t_{ln}] \psi_{mn}.
\end{eqnarray}
Expanding Equation~\eqref{equation2detail} leads to: 
\begin{eqnarray}
\Gamma_{kl} &=& |S_{\mathcal{M}}|^2 + |S_{\mathcal{N}}|^2 + |S_{\mathcal{M},\mathcal{N}}|^2 \nonumber \\
&+& e^{2i(\theta_{\mathcal{M}} - \theta_{\mathcal{N}})} S_{\mathcal{M}} S_{\mathcal{N}}^*
+ e^{-2i(\theta_{\mathcal{M}} - \theta_{\mathcal{N}})} S_{\mathcal{M}}^* S_{\mathcal{N}}  \label{EqX}  \\
&+& e^{i(\theta_{\mathcal{M}} - \theta_{\mathcal{N}})} S_{\mathcal{M}} S_{\mathcal{M},\mathcal{N}}^*
+ e^{-i(\theta_{\mathcal{M}} - \theta_{\mathcal{N}})} S_{\mathcal{M}}^* S_{\mathcal{M},\mathcal{N}}  \nonumber \\
&+& e^{i(\theta_{\mathcal{N}} - \theta_{\mathcal{M}})} S_{\mathcal{N}} S_{\mathcal{M},\mathcal{N}}^*
+ e^{-i(\theta_{\mathcal{N}} - \theta_{\mathcal{M}})} S_{\mathcal{N}}^* S_{\mathcal{M},\mathcal{N}}.  \nonumber
\end{eqnarray}
By identifying the coefficients of Equation~\eqref{EqX} with those of Equation~\eqref{generalInter}, we find:
\begin{eqnarray}
    \theta &=& \theta_\mathcal{M}-\theta_\mathcal{N} \\
    A &=& 2 |S_{\mathcal{M}} S_{\mathcal{N}}^*| \\
    \theta_A &=& \arg(S_{\mathcal{M}} S_{\mathcal{N}}^*) \\
    B &=& 2 |S_{\mathcal{M}} S_{\mathcal{M},\mathcal{N}}^* + S_{\mathcal{N}} S_{\mathcal{M},\mathcal{N}}^*| \\
    \theta_B &=& \arg(S_{\mathcal{M}} S_{\mathcal{M},\mathcal{N}}^* + S_{\mathcal{N}} S_{\mathcal{M},\mathcal{N}}^*) \\
    C &=& |S_{\mathcal{M}}|^2 + |S_{\mathcal{N}}|^2 + |S_{\mathcal{M},\mathcal{N}}|^2.
\end{eqnarray} \\
\\
\noindent \textbf{Determination of the optimal phase.}
Determining the value of $\theta$ that maximizes $\Gamma$ at each optimization step is done in two stages. First, the coefficients $A$, $B$, $C$, $\theta_A$, and $\theta_B$ are determined by measuring $\Gamma$ for six phase values $\{0, \pi/4, \pi/2, \pi, 3\pi/2, 5\pi/4\}$ as follows:  
\begin{eqnarray}
   B e^{i \theta_B} &=& \frac{\Gamma(0)-\Gamma(\pi)}{2} + i \frac{\Gamma({3 \pi/2})-\Gamma({\pi/2})}{2}  \\ 
   C &=& \frac{\Gamma(0)+\Gamma(\pi)+\Gamma({3 \pi/2})+\Gamma({\pi/2})}{4} \\ 
A e^{i \theta_A} &=& \frac{\Gamma(0)+\Gamma(\pi)-2C}{2} + i \frac{\Gamma({\pi/4})+\Gamma({5 \pi/4})-2C}{2}
\end{eqnarray}  
To reduce uncertainty due to measurement noise, one can also measure more than six points and fit them to the model of Equation~\eqref{generalInter} by minimizing the sum of squared errors. Once the coefficients are known, the optimal value of $\theta$ is determined analytically by solving a fourth-degree polynomial equation or using a numerical approach. More details can be found in Supplementary Section III.\\
\\
{\noindent \textbf{Modeling and characterization of the two-photon entangled state.}} 
{In the SLM plane, the two-photon wavefunction $\psi$ is modeled by a double Gaussian distribution~\cite{schneeloch_introduction_2016,fedorov_gaussian_2009}:
\begin{equation}
    \psi(\vec{r_1},\vec{r_2})  =A \, \exp{-\frac{|\vec{r_1}-\vec{r_2}|^2}{4 \sigma_{\vec{r}}^2}} \exp{ -\frac{|\vec{r_1}+\vec{r_2}|^2 \sigma_{\vec{k}}^2}{4}},
\end{equation}
where $A$ is a normalization factor, $\vec{r_1}$ and $\vec{r_2}$ are the idler and signal photons transverse positions in the SLM plane, and $\sigma_\vec{r}$ and $\sigma_\vec{k}$ are the position and momentum correlation widths, respectively. To determine their values, the SPAD camera was replaced by an Electron Multiplied Charge Coupled Device camera (EMCCD) to have a better pixel resolution, and the method detailed in Refs.~\cite{moreau_realization_2012,edgar_imaging_2012} was used. The values $\sigma_\vec{r} = 2.9 \times 10^{-5}\,m$ and $\sigma_\vec{k} = 8.0 \times 10^2\,m^{-1}$ were found. See Supplementary Section V for more details }\\
\\
\noindent \textbf{Measurement of $\Gamma$ and sum-coordinate projection $\Gamma^+$.} The spatially resolved second-order correlation function $\Gamma_{kl}$, where $k$ and $l$ are two pixels of the SPAD camera centred at transverse positions $(x_k,y_k)$ and $(x_l,y_l)$, is measured by acquiring a set of $P+1$ frames $\{I^{(p)} \} _{p \in  [\![ 1,P+1]\!]}$ using a fixed exposure time and processing them using the formula~\cite{defienne_general_2018,ndagano_imaging_2020}: 
\begin{equation}
\Gamma_{kl} = \frac{1}{P} \sum_{p=1}^P \left[  I^{(p)}_{k}I^{(p)}_{l} - I^{(p)}_{k} I^{(p+1)}_{l} \right].
\label{sumequ}
\end{equation}
Because the SPAD camera does not resolve the number of photons, photon coincidences at the same pixel (i.e. coefficients $\Gamma_{kk}$) cannot be measured, and are therefore set to zero. In addition, correlation values between neighboring pixels suffer from cross-talk and must be corrected, as detailed in Ref.~\cite{courme_manipulation_2023}. The sum-coordinate projection of \( \Gamma \) is then calculated using the following formula:
\begin{equation}
\label{equ3s}
\Gamma^+_{T} = \sum_{k=1}^{M} \Gamma_{T-k,k},
\end{equation}
where $T-k$ denotes cameras pixels of transverse position $(x_T-x_k,y_T-y_k)$. {Using a continuous-variable formalism, this definition becomes:
\begin{equation}
\Gamma^+(\vec{r_T})  = \int \Gamma(\vec{r_T}-\vec{r},\vec{r}) d \vec{r},
\end{equation}
where $\vec{r_T} = (x_T,y_T)$ is the transverse position of the target.}\\
\\
\noindent \textbf{Optimization target.} {In our experiment, we chose to use $\Gamma^+_T$ as the optimization target instead of $\Gamma_{kl}$ directly. This choice was motivated by practical considerations: the SNR is higher for $\Gamma^+_T$ than for $\Gamma_{kl}$. Indeed, Equation~\eqref{equ3s} shows that $\Gamma^+_T$ results from a sum of many $\Gamma_{kl}$ terms. If each term is noisy (i.e. $\Gamma_{kl} + \epsilon_{kl}$, with independent noise $\epsilon_{kl}$), then the variance of the resulting noise on $\Gamma^+_T$ is proportional to the number of nonzero terms in the sum. Since the signal values add directly, the overall SNR is higher on average by the square root of the number of terms.
Furthermore, it is important to note that the form of Equation~\eqref{generalInter}, originally derived for $\Gamma_{kl}$, is preserved under summation. The resulting function $\Gamma^+_T$ can thus also be expressed as $A \cos(2\theta + \theta_A) + B \cos(\theta + \theta_B) + C$, where the different parameters are combinations of those associated with the individual terms in the sum.
This choice of target and experimental configuration makes the experiment feasible with current detection technologies while maintaining the general applicability of our approach. Additional details are provided in Supplementary Section VI.}\\
\\
{\noindent \textbf{Construction of the non-entangled states.} We describe here the two non-entangled states used to compare with the entangled-state non-classical optimization. The first non-entangled state chosen is a pure separable state, defined by its wavefunction $\psi(\vec{r_1}, \vec{r_2}) = \phi(\vec{r_1}) \chi(\vec{r_2})$. To reproduce the same sum-coordinate projection $\Gamma^+$ as that of the entangled state measured on the camera without the scattering medium i.e. a Gaussian with width proportional to $\sigma_{\vec{k}}$ shown in Figure~\ref{fig 3}a, the individual wavefunctions in the SLM plane are chosen as:
\begin{equation}
    \phi(\vec{r}) = \chi(\vec{r}) = A' \exp{-\frac{|\vec{r}|^2 \sigma_\vec{k}^2}{8}},
\end{equation}
where $A$ is a normalization constant.\\
The second non-entangled state chosen is a mixed separable state, denoted $\rho$. In this case, one can choose the weights $p_j$ and the wavefunctions $\phi^j$ and $\chi^j$ to exactly reproduce the full second-order correlation function $\Gamma$ measured in the camera plane for the entangled state without the medium - and consequently also $\Gamma^+$.
For that, $\rho$ is defined using continuous-variable formalism as $\rho = \iint p_\vec{q} \ket{\psi^\vec{q}} \bra{\psi^\vec{q}} d\vec{q}$, where:
\begin{align}
\label{modelNE1} \phi^\vec{q}(\vec{r_1}) &= \exp{- \frac{|\vec{r_1}|^2 \sigma_{\vec{k}}^2}{4(1 + \sigma_{\vec{r}}^2 \sigma_{\vec{k}}^2)}} e^{-i \vec{q} \vec{r_1}} \\
\label{modelNE2} \chi^\vec{q}(\vec{r_2}) &= e^{-i \vec{q} \vec{r_2}} \\ 
\label{modelNE3} p_{\vec{q}} &= D, 
\end{align}
where $\psi^\vec{q}(\vec{r_1}) \chi^\vec{q}(\vec{r_2})$ is the (factorizable) two-photon function of the state $\ket{\psi^\vec{q}}$, and $D$ is a constant chosen to ensure proper normalization of the state. More details about these states and the non-classical optimization performed with them are provided in Supplementary Sections XI and XIII.}\\
\\
\noindent \textbf{Details on the simulations.} The simulations presented in Figure~\ref{fig 5} and Supplementary Figures~S13e–h were performed by propagating the two-photon state through the optical system using matrix multiplication in MATLAB. In Supplementary Figures~S13e–h, the scattering matrix was generated from a random phase mask followed by a Fourier transform, producing a disordered medium with statistical properties similar to those observed experimentally.
The state $\rho$ was generated according to the model defined in Equations~\eqref{modelNE1},~\eqref{modelNE2} and~\eqref{modelNE3}, using the experimentally measured values of $\sigma_{\vec{k}}$ and $\sigma_{\vec{x}}$ to closely match the characteristics of the entangled state and the optical setup.
In contrast, for Figure~\ref{fig 5}, the scattering matrix was measured experimentally using the method described in Ref.~\cite{popoff_measuring_2010}. The spatial correlation width in the SLM plane, $\sigma_{\vec{r}}$, was varied from $1.6\,\mu\text{m}$ to $1250\,\mu\text{m}$, while $\sigma_{\vec{k}}$ was held constant. The Schmidt number $K$ was then calculated using the expression $K = \frac{1}{4} \left( \sigma_{\vec{r}} \sigma_{\vec{k}} + (\sigma_{\vec{r}} \sigma_{\vec{k}})^{-1} \right)^2$~\cite{schneeloch_introduction_2016,fedorov_gaussian_2009}. Further details on the simulation methods are provided in Supplementary Sections II and IX.\\
\\
{\noindent \textbf{Energy minimization of a multi-spin Hamiltonian.} To use our system as an optical simulator, we associate each input mode $n$ of the SLM (macro-pixel) with a spin variable $\sigma_n$. The value of this spin, either $+1$ or $-1$, is defined by the square of the phase shift applied to the input mode, which can therefore take physical values of $0$ or $\pi/2$. After propagation through the scattering medium, the correlation value $\Gamma^+_T$ depends on the phase values applied by all the macro-pixels on the SLM. This value is then identified with the energy of a multi-spin interaction Hamiltonian of the following form:
\begin{eqnarray}
    H(\vec{\sigma}) &=& - \frac{1}{2} \sum_{nm} J_{nm}^T \sigma_n \sigma_m- \frac{1}{2} \sum_{nml} \Lambda_{nml}^T \sigma_n \sigma_m \sigma_l   \nonumber \\
    &-&   \frac{1}{2} \sum_{nmlp} Q_{nmlp}^T \sigma_n \sigma_m \sigma_l \sigma_p - \frac{1}{2} \sum_{n} K_{n}^T \sigma_n, 
    \label{Isineq}
\end{eqnarray}
where the coefficients $K_n^T$, $J_{nm}^T$, $\Lambda_{nml}^T$, and $Q_{nmlp}^T$ depend on the transmission matrix elements, the input two-photon wavefunction, the macropixel size of the SLM, and the optimization target $T$. Notably, this Hamiltonian includes three-spin and four-spin interaction terms. To find its ground state, we perform a non-classical optimization on $\Gamma^+_T$, constrained to phase values of $0$ or $\pi/2$ on the SLM macropixels. Figure S12 of the Supplementary Document shows an example of such an optimization implemented with $8 \times 8 = 64$ spins, which converges toward a minimum energy configuration. The resulting SLM pattern - i.e. the set of $8 \times 8$ spin values - represents the ground state of the system, found with an estimated accuracy of $25\%$ due to noise limitations. Further details on this optical simulation, including the derivation of Equation~\eqref{Isineq} and the full expressions of the parameters $K_n^T$, $J_{nm}^T$, $\Lambda_{nml}^T$, and $Q_{nmlp}^T$, are provided in Supplementary Section XII.
}  \\
\\

\bibliography{Biblio2}

\clearpage 

\onecolumngrid

\begin{center}
    {\large \textbf{Supplementary Materials: Non-classical optimization through complex media}}\\[10pt]
    Baptiste Courme$^{1,2}$, Chloé Vernière$^{1}$, Malo Joly$^{2}$, Daniele Faccio$^{3}$, Sylvain Gigan$^{2}$, Hugo Defienne$^{1}$\\[5pt]
    \textit{
    $^1$Sorbonne Université, CNRS, Institut des NanoSciences de Paris, INSP, F-75005 Paris, France\\
    $^2$Laboratoire Kastler Brossel, ENS-Universite PSL, CNRS, Sorbonne Universite, College de France, 24 rue Lhomond, 75005 Paris, France\\
    $^3$School of Physics and Astronomy, University of Glasgow, Glasgow G12 8QQ, UK
    }
\end{center}

\vspace{20pt} % Add some space before content

\setcounter{equation}{0} % Reset equation numbering
\renewcommand{\theequation}{B\arabic{equation}} % Number equations as S1, S2, ...
\renewcommand{\thefigure}{S\arabic{figure}}
\setcounter{section}{0}
\setcounter{figure}{0}

\section{Classical optimization}

In this section, we investigate the solutions that optimize the output intensity at a given spatial mode when a classical coherent state of light is shaped at the input of the scattering medium.

\subsection{Demonstration}

Focusing classical coherent light through a scattering medium by iterative optimization is described in Ref.~\cite{vellekoop_focusing_2007}. In this case, one can show that the optimal solution is unique and that there are no local maxima. To demonstrate this, let us consider the matrix equation describing the propagation of classical coherent light through a scattering medium:  
\begin{equation}
    E^{(o)} = T E^{(i)},
    \label{clamat}
\end{equation}
where \( T \) is the scattering matrix of the medium, \( E^{(o)} \) is a vector representing the output field discretized over the output spatial modes, and \( E^{(i)} \) is a vector representing the input field discretized over the input spatial modes. The intensity at an arbitrary output mode \( k \) is given by  
\begin{equation}
    I_k^{(o)} = \left| \sum_{l=1}^N E_l t_{kl} e^{i (\phi_{kl} + \alpha_l + \theta_l)} \right|^2,
    \label{Iten}
\end{equation}
where \( t_{kl} e^{i \phi_{kl}} \) is the matrix coefficient linking input mode \( l \) to output mode \( k \), \( E_l e^{i \alpha_l} \) is the field incident on mode \( l \) of the spatial light modulator (SLM), \( \theta_l \) is the phase implemented by the SLM at input mode \( l \), and \( N \) is the total number of input modes. To simplify calculations without loss of generality, we redefine the quantities in Eq.~\eqref{Iten} as follows: \( |E_l t_{kl}| \rightarrow a_l \) and \( \phi_{kl} + \alpha_l + \theta_l \rightarrow \theta_l \). In the following, we study the critical points of \( I^{(o)} \) as a function of the variables \( \{\theta_l\}_{l \in \llbracket 1,N\rrbracket} \). By definition, critical points are found by setting the gradient to zero:  
\begin{equation}
    \forall l \in \llbracket 1,N\rrbracket, \quad \frac{\partial I^{(o)}}{\partial \theta_l} = 0.
    \label{grad1}
\end{equation}
Expanding Equation~\eqref{grad1}, we obtain another set of equations:  
\begin{equation}
    \forall l \in \llbracket 1,N\rrbracket, \quad \sum_{p \neq l} a_l a_p \sin(\theta_l - \theta_p) = 0.
    \label{grad2}
\end{equation}
To solve Equation~\eqref{grad2}, we combine the different equations and show that it is necessary for each term to cancel individually i.e. \( \forall (l, p) \in \llbracket 1,N\rrbracket^2, \sin(\theta_l - \theta_p) = 0 \). If we fix \( \theta_N = 0 \) as the reference phase, the critical points of \( I^{(o)} \) are thus given by:  
\begin{equation}
    (\theta_1, \theta_2, \ldots, \theta_{N-1}, \theta_N) = (0 \pm \pi, 0 \pm \pi, \ldots, 0 \pm \pi, 0).
    \label{grad4}
\end{equation}
By analyzing the variations of \( I^{(o)} \) around these critical points, we find that only one critical point is a maximum:  
\begin{equation}
    (\theta_1, \theta_2, \ldots, \theta_{N-1}, \theta_N) = (0, 0, \ldots, 0, 0).
\end{equation}
By performing the inverse change of variables \(\theta_l \rightarrow \phi_{kl} + \alpha_l + \theta_l \), we have:
\begin{equation}
    (\theta_1, \theta_2, \ldots, \theta_{N-1}, \theta_N) = (-\phi_{k1} - \alpha_1, -\phi_{k2} - \alpha_2, \ldots, -\phi_{k{N-1}} - \alpha_{N-1}, -\phi_{k{N}} - \alpha_{N}).
\end{equation}
All other critical points correspond to either minima or saddle points. Hence, there is a unique maximum, which is necessarily the global maximum.

\subsection{Simulations}
To illustrate this, let us consider a simple situation described in Figure~\ref{FigSM1}a. At the input, we consider three spatial modes whose phases $\theta_1$, $\theta_2$, and $\theta_3$ can be independently adjusted. At the output, the intensity is measured at a single spatial mode. The parameter space has an effective dimension of $2$, as the phases are defined up to a global constant i.e. $\theta_1 + \theta_2 + \theta_3 = \text{constant}$.

\begin{figure} [h!]
    \centering
    \includegraphics[width = 0.7\textwidth]{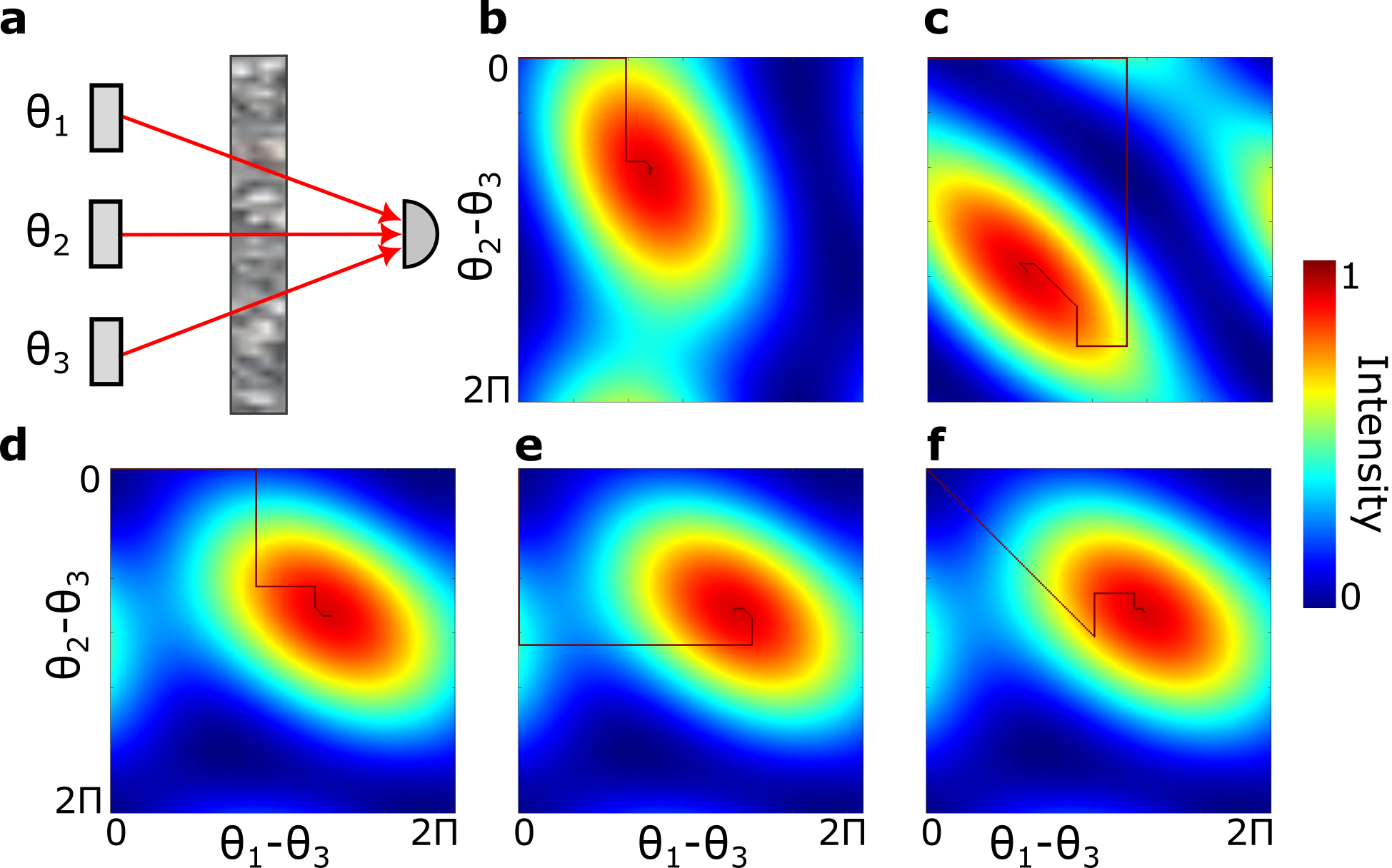}
    \caption{\justifying \textbf{Simulations of classical optimization with three input modes}. \textbf{a,} Classical coherent light propagates through a scattering medium. Three input spatial modes with controllable phases $\theta_1$, $\theta_2$, and $\theta_3$, and one output spatial mode where the intensity is measured, are considered. The transmission matrix $T$ is a $1 \times 3$ matrix composed of random complex coefficients generated as \( T = \texttt{rand}(1,3) \cdot \exp(i \cdot \texttt{rand}(1,3) \cdot 2\pi) \), where \texttt{rand} is a MATLAB function. \textbf{b-d,} Intensity values \( I^{(o)}(\theta_1, \theta_2) \) calculated for three different scattering matrices. The red line corresponds to the iterative optimization path in the parameter space using a partitioning algorithm. \textbf{d-f,} Three distinct iterative optimizations performed for the same scattering matrix, all converging towards the same unique maximum. In each optimization simulation, white noise with a variance of approximately \( 1\% \) of the mean intensity was added to mimic experimental noise.} 
    \label{FigSM1}
\end{figure}

Figures~\ref{FigSM1}b-d show the intensity values at the output mode as a function of the parameters $\theta_1 - \theta_3$ and $\theta_2 - \theta_3$ for different scattering matrices. The scattering matrices $T$ are composed of $1 \times 3$ random complex coefficients. In the simulation, the output intensity value $I^{(o)}(\theta_1,\theta_2)$ for the given phase values $\theta_1$ and $\theta_2$ is obtained by matrix multiplication:
\begin{equation}
    I^{(o)}(\theta_1,\theta_2) = \left |T 
    \begin{bmatrix}
           e^{i \theta_1} \\
           e^{i \theta_2} \\
           1
    \end{bmatrix} \right|^2.
\end{equation}
Here, we calculate \( I^{(o)}(\theta_1, \theta_2) \) for 1000 values of \(\theta_1\) and \(\theta_2\), each ranging from \(0\) to \(2\pi\).  
The red line in each figure shows the path taken by the iterative optimization process. A partitioning algorithm is used to perform the optimization. At each step, \(\theta_1\), \(\theta_2\), or \(\theta_3\) are randomly selected and scanned over the range \(0\) to \(2\pi\). The phase value that maximizes the output intensity is then fixed before proceeding to the next step. We observe that there are no local maxima and that the algorithm always converges to the unique optimal solution. In addition, Figures~\ref{FigSM1}d-f show three different optimization process performed for the same scattering matrix. While the optimization paths are different, we observe that they all converge to the global maximum.

\subsection{Influence of the input field}
Here, the shape of the input field to the SLM does not affect the optimization problem: there will still be a unique maximum and no local maxima. Indeed, if we consider an arbitrary input field denoted as  
\[
E^{(i)} = \begin{bmatrix}
    E^{(i)}_1 \\
    E^{(i)}_2 \\
    E^{(i)}_3
\end{bmatrix},
\]  
the propagation equation can always be rewritten as:  
\[
I^{(o)}(\theta_1, \theta_2) = \left| T 
\begin{bmatrix}
    e^{i \theta_1} & 0 & 0 \\
    0 & e^{i \theta_2} & 0 \\
    0 & 0 & 1
\end{bmatrix} E^{(i)} \right|^2 
= \left| T' 
\begin{bmatrix}
    e^{i \theta_1} \\
    e^{i \theta_2} \\
    1
\end{bmatrix} \right|^2,
\]  
where  
\[
T' = T  
\begin{bmatrix}
    E^{(i)}_1 & 0 & 0 \\
    0 & E^{(i)}_2 & 0 \\
    0 & 0 & E^{(i)}_3
\end{bmatrix}.
\]  
This brings us back to a classical optimization problem, but with a new matrix \(T'\). 

\section{Optimization of a two-photon state}
\label{sectionII}
In this section, we investigate the space of solutions that maximize correlations in the case of a two-photon state propagating through a scattering medium.

\subsection{Reduction of the parameter space}

Equation (1) in the manuscript corresponds to one row of the matrix equation describing the propagation of a two-photon state through a scattering medium:  
\begin{equation}  
    \psi^{(o)} = T \psi^{(i)} T^t,  
\end{equation}  
where \( T \) is the scattering matrix and  
\[
\psi^{(i)} =  
\begin{bmatrix}  
   \psi^{(i)}_{11} e^{i \theta_{11}} & \ldots &  \ldots \\  
    \ldots & \psi^{(i)}_{kl} e^{i \theta_{kl}} & \ldots \\  
    \ldots & \ldots & \psi^{(i)}_{NN}e^{i \theta_{NN}}  
\end{bmatrix},  
\]  where \( N \) is the total number of input modes, \( \theta_{kl} \) are phase terms and \( \psi^{(i)}_{kl} \) represents the two-photon state coefficients. As seen in Equation (1) of the manuscript, in an ideal case, all \( N^2 \) phase terms \( \theta_{kl} \) could be adjusted independently to maximize the correlation value between any pair of output modes, and the optimization problem would then be equivalent to the classical case. However, when using a single SLM, only \( N \) phase terms $\theta_k$ corresponding to the \( N \) input modes can be controlled, leading to \( \theta_{kl} = \theta_k + \theta_l \). In this scenario, the effective accessible parameter space is reduced and of smaller dimension.

\begin{figure} [h!]  
    \centering  
    \includegraphics[width=0.7\textwidth]{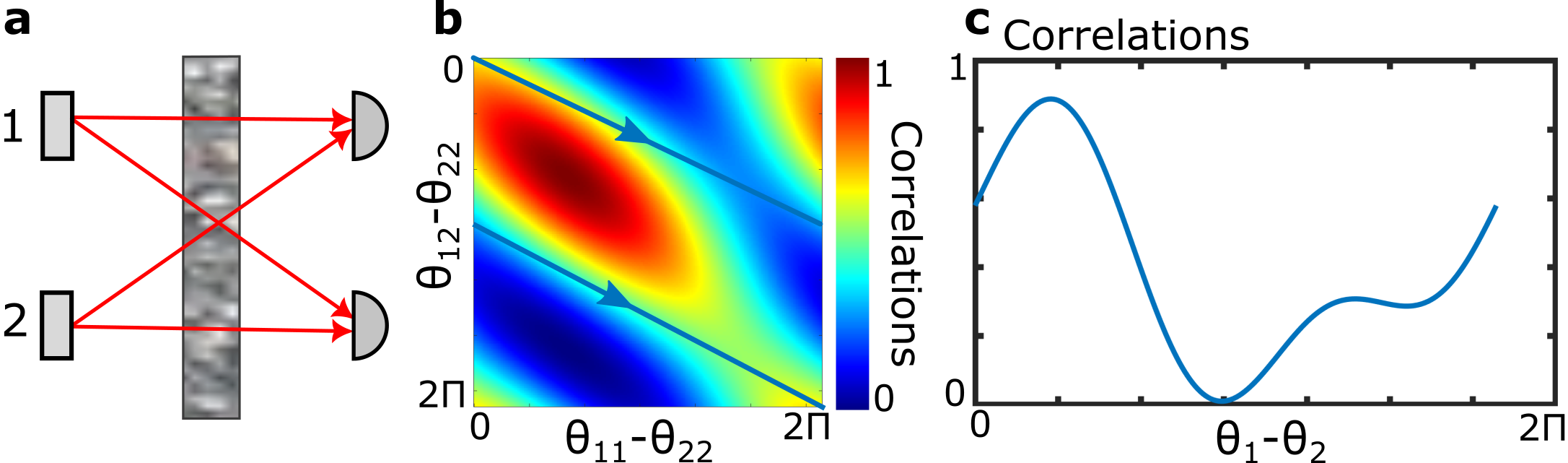}  
    \caption{\justifying \textbf{Simulations of the parameter space for two-photon state propagation and shaping through a scattering medium}.  
    \textbf{a,} A two-photon state propagates through a scattering medium. Two input spatial modes and two output spatial modes, between which the correlations are measured, are considered. The transmission matrix \( T \) is a \( 2 \times 2 \) matrix with random complex coefficients generated as \( T = \texttt{rand}(2) \cdot \exp(i \cdot \texttt{rand}(2) \cdot 2\pi) \), where \texttt{rand} is a MATLAB function. 
    \textbf{b,} Correlation values \( \Gamma^{(o)}(\theta_{11}, \theta_{12}) \) are calculated assuming full control over the phase terms \( \theta_{11} \), \( \theta_{12} \), and \( \theta_{22} \) (with \( \theta_{11} + \theta_{12} + \theta_{22} = \text{constant}\) ). The blue line represents the reduced parameter space under the experimental constraint that only two phases, \( \theta_1 \) and \( \theta_2 \), can be arbitrarily controlled (with \( \theta_{1} + \theta_{2} = \text{constant} \)). The relationship between the phase terms is \( \theta_{11} = 2\theta_1 \), \( \theta_{12} = \theta_1 + \theta_2 \), and \( \theta_{22} = 2\theta_2 \).  
    \textbf{c,} Correlation values as a function of \( \theta_1 - \theta_2 \) (i.e. corresponding to the blue line in \textbf{(b)}), revealing that local maxima can exist, and that the global maximum here does not necessarily reach the global maximum of the full parameter space.}  
    \label{FigSM2}  
\end{figure}  

For example, Figure~\ref{FigSM2}a illustrates a simple case with two input modes and two output modes. If all the phases (i.e., \( \theta_{11} \), \( \theta_{12} \), and \( \theta_{22} \)) can be controlled, the parameter space is two-dimensional, and the correlation map in Figure~\ref{FigSM2}b shows a unique global maximum with no local maxima, similar to the classical case. However, in practice, only \( \theta_1 \) and \( \theta_2 \) can be controlled, reducing the parameter space to one dimension. As shown in Figure~\ref{FigSM2}c, the variation of output correlations is a curve that does not necessarily reach the global maximum of the full parameter space (Fig~\ref{FigSM2}b). It can exhibit local maxima and potentially multiple solutions.  

\subsection{Influence of the input two-photon field}

As we have seen in the previous section, the parameter space accessible in practice is reduced. In particular, the properties of this space depends on the form of the input state incident on the SLM. To illustrate this, Figure~\ref{FigSM3} presents simulations of a simple case with 3 input modes and 2 output modes. In these simulations, the same scattering matrix, composed of \(3 \times 2\) random complex coefficients, is generated. Parameter spaces are then calculated, and an iterative optimization is performed for three different forms of input states:

\begin{itemize}
    \item Figure~\ref{FigSM3}b shows the output correlation rates in function of \(\theta_1-\theta_3\) and \(\theta_2-\theta_3\) for an input state of the form \(|\Psi\rangle = 1/\sqrt{3} [a_1^{\dagger 2} + a_2^{\dagger 2} + a_3^{\dagger 2}] |0\rangle\). In this case, we observe the presence of genuine multiple global maxima (i.e. identical values) of the output correlation, between which the optimization process can oscillate. This state is very similar to the one incident on the SLM when using the experimental configuration presented in Figure 2 of the manuscript.
    \item Figure~\ref{FigSM3}c shows the output correlation rates in function of \(\theta_1-\theta_3\) and \(\theta_2-\theta_3\) for an input state of the form \(|\Psi\rangle = [0.577 a_1^{\dagger 2} + 0.577 a_2^{\dagger 2} + 0.577 a_3^{\dagger 2} + 0.002 a_1^\dagger a_2^\dagger + 0.002 a_2^\dagger a_3^\dagger] |0\rangle\). In this case, we observe the presence of global and local maxima with very similar correlation values. In practice, the presence of experimental noise makes them indistinguishable, and the optimization process can also oscillate between these different attractors. 
    \item Figure~\ref{FigSM3}d show the output correlation rates in function of \(\theta_1-\theta_3\) and \(\theta_2-\theta_3\) for an input state of the form \(|\Psi\rangle = [0.35 a_1^{\dagger 2} + 0.35 a_2^{\dagger 2} + 0.35 a_3^{\dagger 2} + 0.16 a_1^\dagger a_2^\dagger + 0.16 a_2^\dagger a_3^\dagger + 0.002 a_1^\dagger a_3^\dagger] |0\rangle\).  Here again, we observe the presence of local maxima with values approaching those of the global maxima. 
    \item Figure~\ref{FigSM3}e and f shows the output correlation rates in function of \(\theta_1-\theta_3\) and \(\theta_2-\theta_3\) for two different input state of the form \(|\Psi\rangle = \mathcal{N}\sum_{k=1}^3 \sum_{l=k}^3 \alpha_{kl} a_k^\dagger a_l^\dagger |0\rangle\), where \(\alpha_{kl}\) are random complex coefficients and \(\mathcal{N}\) is a normalization factor. We clearly observe that the parameter space strongly depends on the form of the input state. In Figure~\ref{FigSM3}e, it presents multiple maxima of similar values between which the optimization can oscillate. In Figure~\ref{FigSM3}e, however, there is only one clear maximum to which the optimization always converges.
\end{itemize}

These simulations highlight the complexity of the reduced parameter space when using a single SLM to shape a two-photon state. In particular, they confirm that local maxima and multiple global maxima can exist, but their presence strongly depends on the form of the two-photon input state shaped by the SLM. In our experimental configuration, the crystal and the SLM are positioned so that the two-photon state incident on the SLM is not random but has a shape of the approximate form: $| \Psi \rangle = \left [ J_0 \sum_{k=1}^N a_k^{\dagger 2} + J_1 \sum_{k=1}^{N-1} a_k^\dagger a_{k+1}^\dagger + + J_2 \sum_{k=1}^{N-2} a_k^\dagger a_{k+2}^\dagger  + \ldots  \right ] |0\rangle$, with $J_n \gg J_{n+1}$. we are therefore operating in a situation close to those simulated in Figures~\ref{FigSM3}b-d, where multiple maxima of similar values seems to always exist.

\begin{figure} [h!]  
    \centering  
    \includegraphics[width=0.7\textwidth]{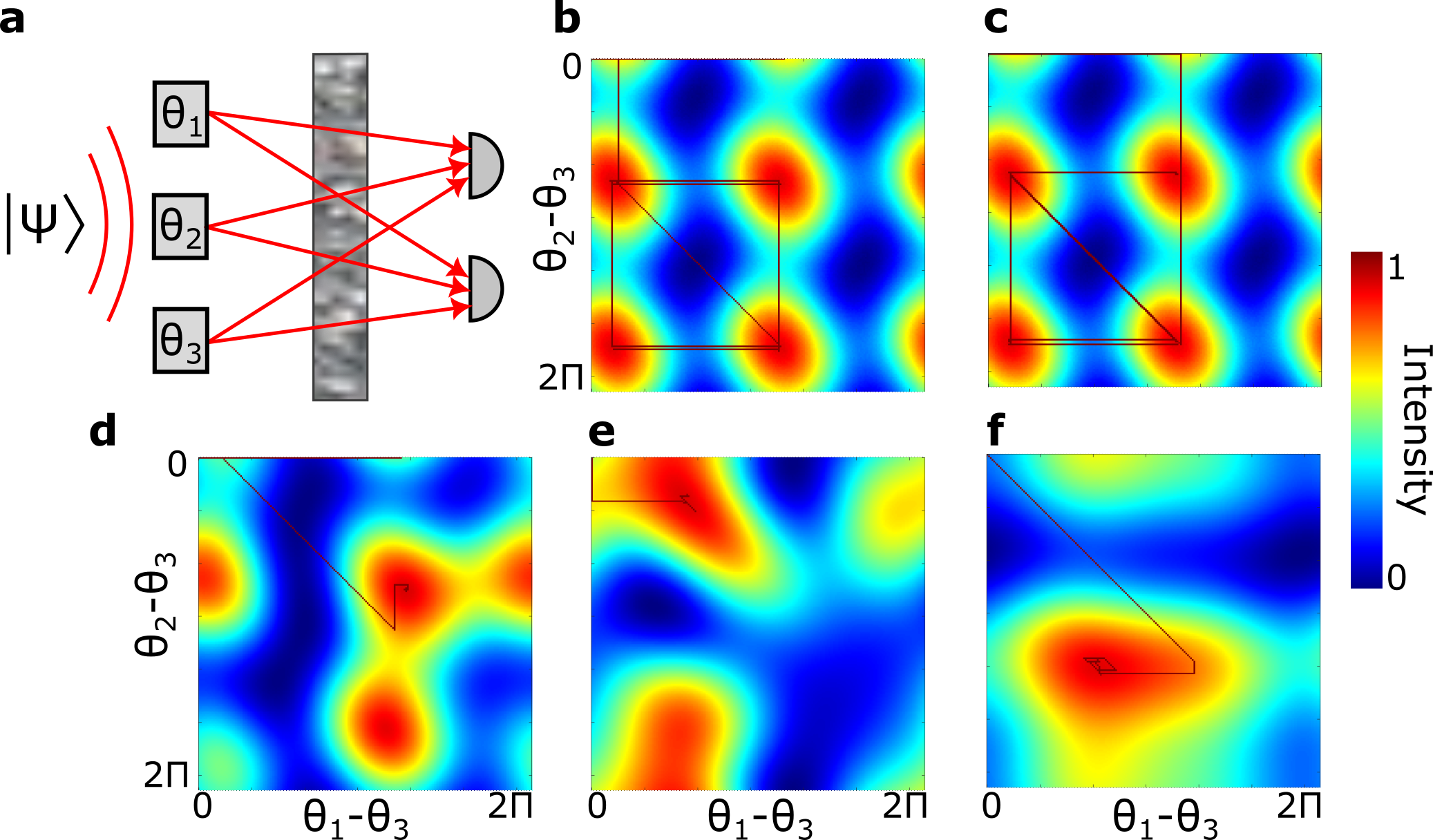}  
    \caption{\justifying \textbf{Simulations of iterative optimizations for different input states propagating through the same scattering medium.} \textbf{a,} A two-photon state propagates through a scattering medium. Three input spatial modes and two output spatial modes, between which the correlations are measured, are considered. The transmission matrix \( T \) is a \( 2 \times 3 \) matrix with random complex coefficients, generated as \( T = \texttt{rand}(2,3) \cdot \exp(i \cdot \texttt{rand}(2,3) \cdot 2\pi) \), where \texttt{rand} is a MATLAB function. Correlation values \( \Gamma^{(o)}(\theta_{11}, \theta_{12}) \) are plotted as a function of \( \theta_1 - \theta_2 \) and \( \theta_1 - \theta_3 \) for different input states incident on the SLM: \textbf{(b)} \(| \Psi\rangle = \mathcal{N} [a_1^{\dagger 2} + a_2^{\dagger 2} + a_3^{\dagger 2}] |0\rangle\), \textbf{(c)} \(|\Psi\rangle = [0.577 a_1^{\dagger 2} + 0.577 a_2^{\dagger 2} + 0.577 a_3^{\dagger 2} + 0.002 a_1^\dagger a_2^\dagger + 0.002 a_2^\dagger a_3^\dagger] |0\rangle\), \textbf{(d)} \(|\Psi\rangle = [0.35 a_1^{\dagger 2} + 0.35 a_2^{\dagger 2} + 0.35 a_3^{\dagger 2} + 0.16 a_1^\dagger a_2^\dagger + 0.16 a_2^\dagger a_3^\dagger + 0.002 a_1^\dagger a_3^\dagger] |0\rangle\), and \textbf{(e-f)} \(|\Psi\rangle =  \mathcal{N}\sum_{k=1}^3 \sum_{l=k}^3 \alpha_{kl} a_k^\dagger a_l^\dagger |0\rangle\) with two different sets of \(\alpha_{kl}\), which are random complex coefficients. \(\mathcal{N}\) is a normalization factor for each state. The red line represents the iterative optimization path in parameter space using a partitioning algorithm. In each optimization simulation, white noise with a variance of approximately \( 1\% \) of the mean correlation value was added to mimic experimental noise.}  
    \label{FigSM3}  
\end{figure}  

\subsection{Influence of the scattering matrix}

Figure~\ref{FigSM4} shows the parameter space for different scattering matrices, in the simple case of 3 input modes and 2 output modes. Interestingly, as shown in Figures~\ref{FigSM4}a and \ref{FigSM4}b, when the input state is close to a maximally entangled state, i.e., \(| \Psi\rangle \approx \mathcal{N} [a_1^{\dagger 2} + a_2^{\dagger 2} + a_3^{\dagger 2}] |0\rangle\), there always exist maxima with very close values. As explained at the end of the previous subsection, in our work the SLM and the crystal are positioned in a configuration that approximately satisfies this condition. When the input state is random, however, Figures~\ref{FigSM4}c and \ref{FigSM4}d confirm that the parameter space can be very different. It may or may not have multiple maxima with similar values, and it can lead to very complex optimization paths. 

\begin{figure} [h!]  
    \centering  
    \includegraphics[width=0.5\textwidth]{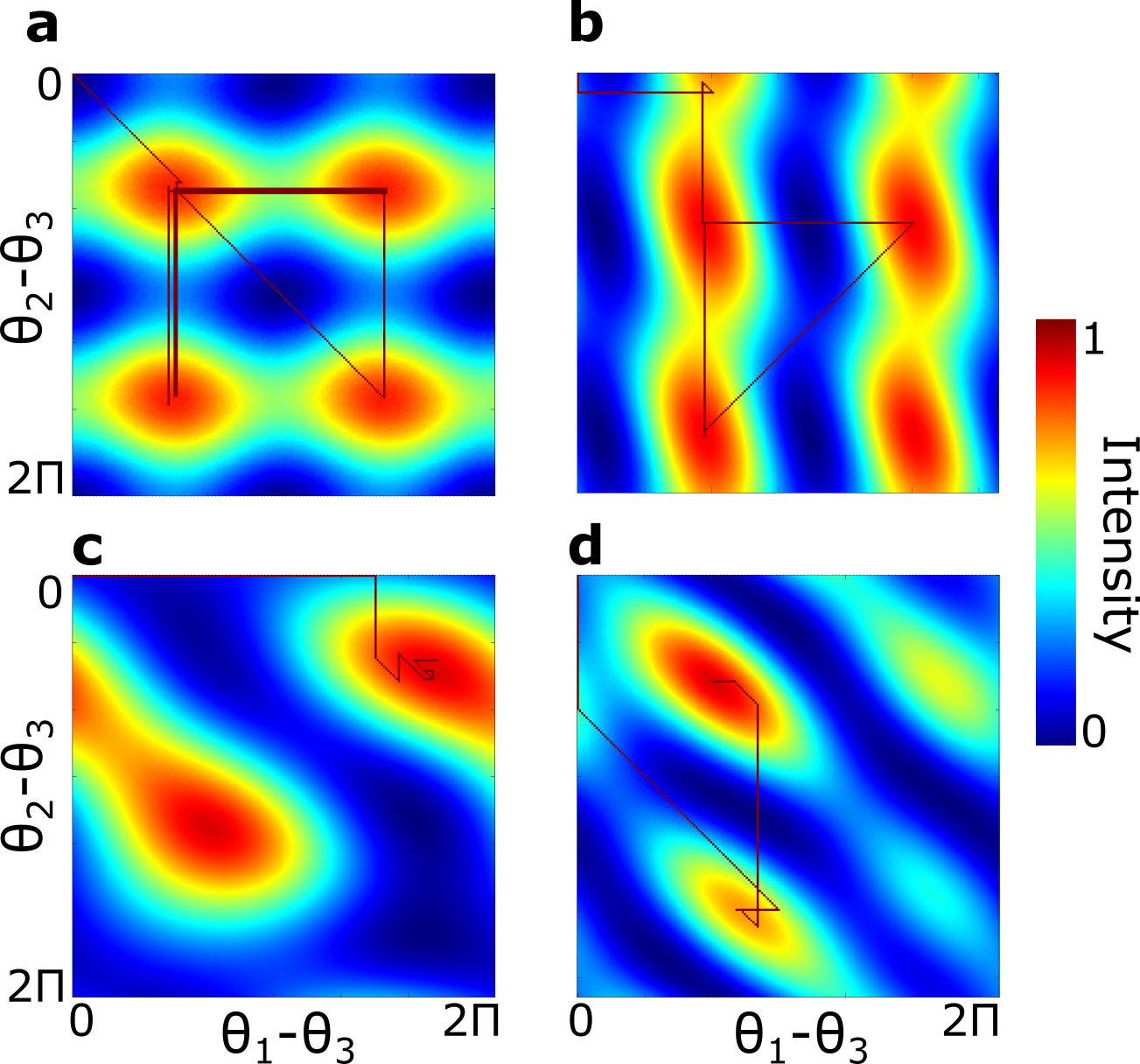}  
    \caption{\justifying \textbf{Simulations of iterative optimizations with an input two-photon state for different scattering matrices}.  
    In this simulation, we also consider the optical arrangement described as in Figure~\ref{FigSM3}a. \textbf{a and b,} Correlation values \( \Gamma^{(o)}(\theta_{11}, \theta_{12}) \) as a function of \( \theta_1 - \theta_2 \) and \( \theta_1 - \theta_3 \) for two different scattering matrices (generated as \( T = \texttt{rand}(2,3) \cdot \exp(i \cdot \texttt{rand}(2,3) \cdot 2\pi) \)) and an input state incident on the SLM of the form \(| \Psi\rangle = \mathcal{N} [a_1^{\dagger 2} + a_2^{\dagger 2} + a_3^{\dagger 2}] |0\rangle\). \textbf{c and b,} Correlation values for two different scattering matrices and an input state incident on the SLM of the form \(|\Psi\rangle = \sum_{k=1}^3 \sum_{l=k}^3 \alpha_{kl} a_k^\dagger a_l^\dagger |0\rangle\), where \(\alpha_{kl}\) are random complex coefficients. The red line represents the iterative optimization path in parameter space using a partitioning algorithm. In each optimization simulation, white noise with a variance of approximately \( 1\% \) of the mean correlation value was added to mimic experimental noise.}  
    \label{FigSM4}  
\end{figure}  

\section{Additional details on finding the optimal phase value}
In this section, we provide an analytical method to find the maximum of the function $\Gamma(\theta) = A \cos(2 \theta + \theta_A) + B \cos(\theta + \theta_B)+C$ described in Equation (2) of the manuscript, where $\theta \in [0,2 \pi [$ is the variable and \(\theta_A \in [0, 2 \pi[\), \(\theta_B \in [0, 2 \pi[\), \(A \geq 0\) and \(B \geq 0\), and \(\theta_B\) are known parameters. These parameters are determined from experimental measurements using Equations (6), (7), and (8) or directly by fitting the data with the model in Equation (2). To find the maximum of $\Gamma(\theta)$, denoted $\theta_{max}$, we first set its derivative to zero: 
\begin{equation}
    \frac{d \Gamma}{ d \theta} = -2 A \sin(2 \theta+\theta_A)-B \sin(\theta+\theta_B) = 0.
    \label{gamm1}
\end{equation}
Then, let's eliminate some trivial solutions. If $A=0$, then it is clear that $\theta_{max} = - \theta_B$. If $B=0$, then there are two solutions: $\theta_{max} = \{ - \theta_A/2,  - \theta_A/2 + \pi\}$. If $\theta_A = 2 \theta_B + \pi$, then $\theta_{max} = - \theta_B + \pi/2$. If $\theta_A = 2 \theta_B$, then $\theta_{max} = - \theta_B$. If $A\neq 0$ and $B \neq 0$, we can rewrite Equation~\eqref{gamm1} as follows: 
\begin{equation}
    \frac{1}{2} \sin(2 x) + D \sin(x + \phi) = 0,
    \label{gamm2}
\end{equation}
with the new variables $x= (2 \theta+\theta_A)/2$, $D=B/(4A)$ and $\phi = \theta_B-\theta_A/2$. Equation~\eqref{gamm2} then expands as
\begin{eqnarray}
    \frac{1}{2} \sin(2 x) + D \sin(s + \phi) &=&  \cos(x) \sin(x) + D \sin(x) \cos(x) + D \cos(\phi) \sin(x) \nonumber \\
    &=& XY+eX+fY = 0,
    \label{gamm3}
\end{eqnarray}
with the new variables $X=\cos(x)$, $Y=\sin(x)$, $e=D \sin(\phi)$ and $f=D \cos(\phi)$. The problem can thus be written in the form of a system of two equations: 
\[
\begin{cases}
    XY + eX + fY = 0 \\
    X^2 + Y^2 = 1.
\end{cases}
\]
Then, we can write: 
\begin{eqnarray}
    X(Y+e) + fY = 0 &\implies& X = \frac{-f Y}{Y+e} \nonumber \\
    &\implies& \frac{f^2 Y^2}{(Y+e)^2} + Y^2 = 1 \nonumber \\
    &\implies& Y^4 + 2eY^3 + [f^2-1+e^2] Y^2 - 2 e Y - e^2 = 0,
    \label{gamm5}
\end{eqnarray}
where the pathological case \(Y = -e\) corresponds to the trivial solutions \(\theta_A = 2 \theta_B\) and \(\theta_A = 2 \theta_B + \pi\), which were already found previously. Finally, Equation~\eqref{gamm5} is a degree-4 equation, which can be solved analytically. Once the different solutions are obtained (up to 4), they must be substituted into Equation~\eqref{gamm1}, and the one yielding the highest value is the (global) maximum. Note that it is also possible to find $\theta_{max}$ using a numerical method. 

\section{Additional details on the experimental setup }
The pump is a collimated continuous-wave laser at $405$ nm (Coherent OBIS-LX) with an output power of $100$ mW and a beam diameter of $0.8\pm 0.1$ mm. The classical source is a superluminescent diode (SLED) centered at 810nm and has a total bandwdith of approximately 20nm (Superlum). BBO crystal has dimensions $0.5 \times 5 \times 5$ mm and is cut for type I SPDC at $405$ nm with a half opening angle of $3$ degrees (Newlight Photonics). The crystal is slightly rotated around horizontal axis to ensure near-collinear phase matching of photons at the output (i.e. ring collapsed into a disk). A $650$ nm-cut-off long-pass filter is used to block pump photons after the crystals, together with a band-pass filter centered at $810 \pm 5$ nm. The SLM has $1080\times1920$ pixels with a pitch of $8\mu$m and uses a liquid crystal on silicon technology  (Holoeye model Pluto-NIR-II). The SLM is tilted and a phase ramp is continuously displayed on it to work with the first order of diffraction. The SPAD camera is the model SPC3 from Micron Photon Device. It has $32 \times 64$ pixel, a pixel pitch of $45 \mu$m and is operated in the snap-running mode with an exposure time of $1 \mu$s. On the same plane as the SPAD, a CCD camera  (uEye) with $1024$ by $1280 $ pixel to perform measurement when the classical source is used. The $4f$ imaging system $f_1-f_2$ in Figure 3.a is represented by two lenses for clarity, but is in reality composed of a series of $4$ lenses with focal lengths $50$ mm - $150$ mm - $100$ mm - $200$ mm. The first and the last lens are positioned at focal distances from the crystal and the SLM, respectively, and the distance between two lenses in a row equals the sum of their focal lengths. The SPDC beam occupies an area of $300$ by $300$ pixels on the SLM plane. Similarly, the second $4f$ imaging system $f_3-f_4$ in Figure 3.a is composed of a series of $4$ consecutive lenses with focal lengths $200$mm - $100$ mm - $75$ mm - $50$ mm arranged as in the previous case. The other lenses have the following focal lengths: $f_5=150$mm. 

\section{Characterization of the two-photon state in the SLM plane}

\subsection{Experimental characterization}

\label{charac}
The two photon state $\psi(\vec{r_1},\vec{r_2})$ produced by a Type-I thin non linear crystal can be expressed analytically using the double Gaussian model~\cite{schneeloch_introduction_2016}: 
\begin{equation}
\label{gaussian}
\psi(\vec{r_1},\vec{r_2})  =A \, \exp{ \frac{-|\vec{r_1}-\vec{r_2}|^2}{4 \sigma_{\vec{r}}^2} } \exp{ \frac{-|\vec{r_1}+\vec{r_2}|^2 \sigma_{\vec{k}}^2} {{4} } },
\end{equation}
where \(\sigma_{\vec{r}}\) and \(\sigma_{\vec{k}}\) are the position and momentum correlation widths, respectively. To determine their values in our experimental setup, we used an EMCCD camera and the method detailed in Refs.~\cite{moreau_realization_2012,edgar_imaging_2012}. 

For \(\sigma_{\vec{r}}\), we added a lens with \(f_6 = 30\) mm between the last lenses \(f_4\) and \(f_5\), setting all lens pairs in a \(4f\) configuration, in order to image the surface of the crystal onto the camera (Figure 2f of the manuscript). Figures~\ref{SM_2}a and b show the intensity image and the minus-coordinate projection of the second-order correlation function, respectively. The width of the peak in Figure~\ref{SM_2}b corresponds to the position correlation width \(\sigma_{\vec{r}}\). It is estimated to be $3$ pixels using a Gaussian fit, as shown in Figure~\ref{SM_2}c. Taking into account that the magnification between the SLM and the camera is $5/3$ and that the camera pixel size is $16$ micrometers, we find that the position correlation width in the SLM plane is \(\sigma_{\vec{r}} \approx 2.9 \cdot10^{-5} \, \text{m}\). 

\(\sigma_{\vec{k}}\) is estimated directly from the width \(\Sigma\) of the intensity image shown in Figure~\ref{SM_2}a, again using a Gaussian fit. Indeed, this width is directly related to \(\sigma_{\vec{k}}\) by the formula \(\sigma_{\vec{k}} = 1/\sqrt{\Sigma^2 - \sigma_{\vec{r}}^2}\)~\cite{schneeloch_introduction_2016}. Taking into account the magnification and the pixel size, we find \(\sigma_{\vec{k}} \approx 8.0 \cdot 10^2 \, \text{m}^{-1}\). Note that it is also possible to measure \(\sigma_{\vec{k}}\) directly by Fourier-imaging the crystal onto the EMCCD camera, as shown in Figures~\ref{SM_2}d-f.

{The Schmidt number, quantifying the degree of entanglement in the state, is evaluated to $K \approx 465$ from the values of $\sigma_{\vec{r}}$ and $\sigma_{\vec{k}}$ using the formula~\cite{schneeloch_introduction_2016}:
\begin{equation}
    K=\frac{1}{4} \left ( \sigma_{\vec{r}} \sigma_{\vec{k}} + \frac{1}{\sigma_{\vec{k}} \sigma_{\vec{r}}} \right )^2.
\end{equation}}

\begin{figure} [h!]
    \centering
    \includegraphics[width = 1\textwidth]{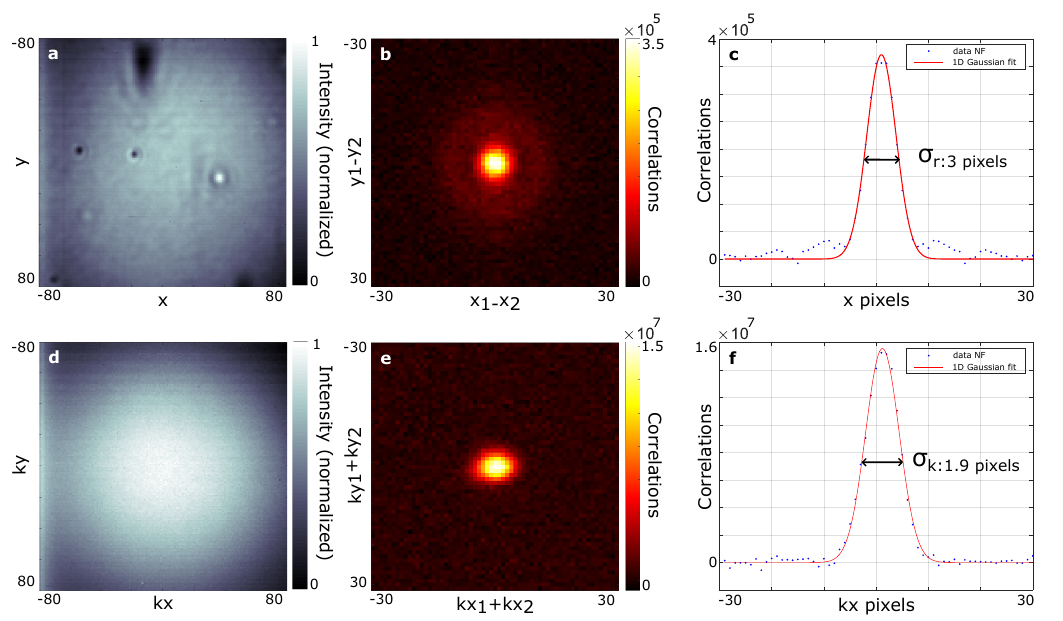}
    \caption{\justifying \textbf{Experimental measurements of position and momentum correlation widths}. \textbf{a and b}, Intensity image and minus-coordinate projection of the second-order correlation function when imaging the surface of the crystal on an EMCCD camera. \textbf{c}, Fitting the peak in the minus-coordinate projection with a Gaussian returns a position correlation width of $3$ pixels. The corresponding value in the SLM plane is calculated by considering the magnification between the SLM and camera planes and the camera pixel size, resulting in \(\sigma_{\vec{r}} \approx 2.9 \cdot10^{-5} \, \text{m}\). Using a Gaussian fit on the intensity image, the total width \(\Sigma \approx 130\) pixels is found, allowing us to determine \(\sigma_{\vec{k}} \approx 8.0 \cdot 10^2 \, \text{m}^{-1}\). \textbf{d-f}, \(\sigma_{\vec{k}}\) can also be estimated directly by Fourier-imaging the crystal on the EMCCD camera, replacing the SPAD camera in the configuration shown in Figure 2f of the manuscript. In this configuration, we measure the intensity image \textbf{(d)} and the sum-coordinate projection of the second-order correlation function \textbf{(e)}. The width of the peak in the latter allows us to determine a width of approximately \(1.9\) pixels using a Gaussian fit \textbf{(f)}. When converted to the SLM plane, the value is consistent with the one directly measured from the intensity image.}  
    \label{SM_2}
\end{figure}

{
\subsection{Propagation to the output plane and sum-coordinate projection}

As shown in Figure 3a of the manuscript, the state described by Equation~\eqref{gaussian} in the SLM plane (input plane) produces strong anti-correlations when it is propagated and detected in the camera plane (output plane). In the absence of scattering medium, the state undergoes a Fourier transform between the input and output planes. For clarify, we describe this transformation by a lens of focal length $f$. The two-photon wavefunction in the camera plane, noted $\psi'$, remains Gaussian and can be written as:
\begin{eqnarray}
\psi'(\vec{r'_1},\vec{r'_2})  &=& \iint \psi(\vec{r_1},\vec{r_2}) e^{i \frac{2 \pi}{\lambda f} \vec{r_1} \vec{r'_1}} e^{i \frac{2 \pi}{\lambda f} \vec{r_2} \vec{r'_2}} d \vec{r_1} d\vec{r_2} \\
&=& A' \, \exp{  \frac{-|\vec{r'_1}-\vec{r'_2}|^2 \sigma_{\vec{r}}^2}{\lambda^2 f^2} } \exp{ \frac{-|\vec{r'_1}+\vec{r'_2}|^2 } {{\lambda^2 f^2 \sigma_{\vec{k}}^2} } },
\end{eqnarray}
where $A'$ is a normalization constant. The corresponding sum-coordinate projection $\Gamma^+$ is then calculated using its definition: 
\begin{eqnarray}
\label{eqGaussianSum}
\Gamma^+(\vec{r_+})  &=& \int \Gamma(\vec{r_+}-\vec{r},\vec{r}) d \vec{r} \\
&=& \int |\psi'(\vec{r_+}-\vec{r},\vec{r})|^2 d \vec{r} \\
&=& B \exp{ \frac{-2|\vec{r_+}|^2 } {{\lambda^2 f^2 \sigma_{\vec{k}}^2} } } \int  \exp{  \frac{-2 |\vec{r_+}-2 \vec{r}|^2 \sigma_{\vec{r}}^2}{\lambda^2 f^2} }  d \vec{r} \\
&=& C \exp{ \frac{-2 |\vec{r_+}|^2 } {{\lambda^2 f^2 \sigma_{\vec{k}}^2} } },
\end{eqnarray}
where $C$ is a normalization constant. As a result, the sum-coordinate projection $\Gamma^+$ observed in the camera plane without the medium is modeled by a Gaussian shape with a width $\sigma_{\vec{k}} \lambda f / \sqrt{2}$, confirming our experimental observation in Figure 3a.}

\section{Validity of the Equation (2) for the optimization target $\Gamma^+_T$}

As demonstrated in the Methods of the manuscript, $\Gamma_{kl}$ varies according to the expression given in Equation (2) of the manuscript when a phase shift $\theta$ is applied between two SLM input modes. By definition, $\Gamma^+_T$ is a sum over many such $\Gamma_{kl}$ terms. Consequently, $\Gamma^+_T$ can be expressed, in very general form, as:
\begin{equation}
    \Gamma^+_T(\theta) = \sum_k A_k \cos(2 \theta + \theta_{A_k}) + B_k \cos(\theta + \theta_{B_k}) + C_k,
\end{equation}
where the parameters $A_k$, $B_k$, $C_k$, $\theta_{A_k}$, and $\theta_{B_k}$ correspond to those of the individual terms in the sum. This follows directly from the observation that:
\begin{eqnarray}
    \Gamma^+_T(\theta) &=& \mathfrak{Re} \left[\sum_k A_k e^{i(2 \theta + \theta_{A_k})} + B_k e^{i(\theta + \theta_{B_k})} \right] + \sum_k C_k \nonumber \\
    &=& \mathfrak{Re} \left[ e^{i 2 \theta } \sum_k A_k e^{i \theta_{A_k}} \right] + \mathfrak{Re} \left[e^{i\theta}\sum_k B_k e^{i \theta_{B_k}} \right] + \sum_k C_k \nonumber \\
    &=& \mathfrak{Re} \left[ e^{i 2 \theta } A e^{i \theta_{A}} \right] + \mathfrak{Re} \left[e^{i\theta} B e^{i \theta_{B}} \right] + C \nonumber \\
    &=& A \cos(2 \theta + \theta_{A}) + B \cos(\theta + \theta_{B}) + C,
\end{eqnarray}
where $C= \sum_k C_k$, $A=\sqrt{|\sum_{k} A_k \cos(\theta_{A_k})|^2+|\sum_{k} A_k \sin(\theta_{A_k})|^2}$, $B=\sqrt{|\sum_{k} B_k \cos(\theta_{B_k})|^2+|\sum_{k} B_k \sin(\theta_{B_k})|^2}$, $\theta_A =\arctan \left( \frac{\sum_k A_k \sin(\theta_{A_k})}{\sum_k A_k \cos(\theta_{A_k})} \right)$ and $\theta_B =\arctan \left( \frac{\sum_k B_k \sin(\theta_{B_k})}{\sum_k B_k \cos(\theta_{B_k})} \right)$.

\section{Influence of the number of controlled input modes on the optimization process}

When focusing through a scattering medium, a key parameter for evaluating the efficiency of the method is the enhancement factor at the end of the process i.e. once a plateau is reached. For classical optimization, this enhancement factor scales linearly with the number of modes controlled by the SLM, as detailed in Ref.~\cite{vellekoop_focusing_2007}. A similar scaling is expected to be observed in the non-classical optimization case, as suggested in Ref.~\cite{lib_real-time_2020}. To confirm this, we perform the non-classical optimization for different numbers of input modes i.e. SLM macropixels. The results for three configurations are shown in Figure~\ref{FigSM10}a: the blue curve corresponds to an active area of $4 \times 4$ macropixels, the red curve (as presented in the main text) corresponds to $8 \times 8$, and the violet curve corresponds to $16 \times 16$. 

In the $4 \times 4$ case, the algorithm reaches a plateau quickly, within just 40 steps ($\sim 1$ day of acquisition). The enhancement factor for this configuration is computed to be \(\eta \approx 3 \). For the $8 \times 8$ case, the plateau was achieved after 200 steps with and enhancement factor computed to be \(\eta \approx5 \) ($\sim 5$ days of acquisition). For the $16 \times 16$ case, even after 400 steps (more than 10 days), the plateau had not been reached. At this point, we decided to halt the experiment. Based on the shape of the optimization curve in this final scenario, we are confident that the plateau, when reached, will be higher than that observed for the $8 \times 8$ case.

While we do not have enough data to determine the precise scaling between the enhancement factor and the number of input modes, these experiments still confirmed that the larger the number of controlled input modes, the higher the enhancement factor.From a practical standpoint, these experiments primarily allowed us to favor the \(8 \times 8\) configuration in all our experiments, as it provides the optimal trade-off between acquisition time and enhancement factor. {Note that the scaling of the enhancement factor with the number of controlled input modes has been derived in a recent theoretical study~\cite{shekel2025fundamentalboundswavefrontshaping}. Both their results and ours support an increase in the enhancement factor with the number of modes}

\begin{figure}[h!]
    \centering
    \includegraphics[width = 1\textwidth]{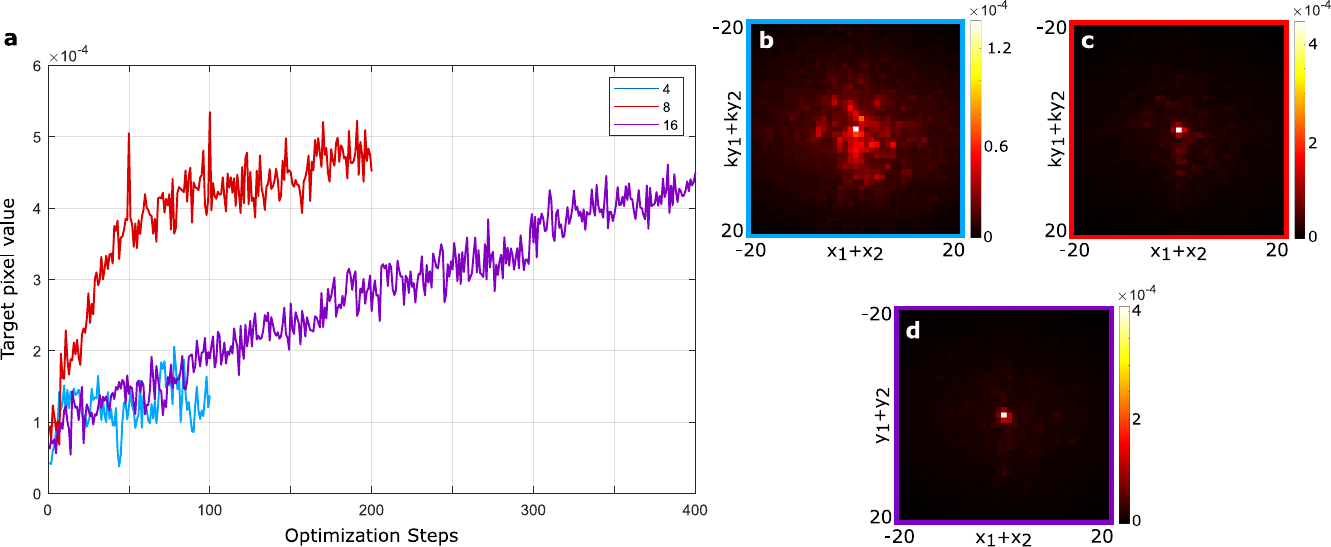}
    \caption{\justifying \textbf{Non-classical optimizations for various number of controlled input modes.} \textbf{a,} Experimentally measured optimization curves for $4 \times 4$ SLM macropixels (blue curve), $8 \times 8$ macropixels (red curve) and $16 \times 16$ macropixels (violet curve). \textbf{b-d,} Sum coordinate projections obtained at the end of the optimization process in each case.}
    \label{FigSM10}
\end{figure}

\section{Improving optimization speed using a Tpxcam sensor}

The practical implementation of the optimization procedure demonstrated in this manuscript is currently limited by its speed, which restricts its use to static media and a relatively small number of controllable modes. Here, we present preliminary results that address these technological limitations. We replaced the SPAD camera with a TpxCam sensor, a single-photon, event-based detector~\cite{nomerotski_imaging_2019,courme_quantifying_2023}. This sensor is capable of detecting photon coincidences within a temporal window of 6~ns and exhibits significantly better quantum efficiency at 810~nm. 

In this experiment, the optimization is performed using a \(16 \times 16\) phase mask. Correlations are measured over a 3-second exposure, and the sum-coordinate projection is computed from the detected events. Each optimization step consists of the application of $10$ different phase masks, as illustrated in Figure~\ref{TPX_speed_up}a, and the full sequence lasts approximately 450 seconds. The additional time beyond the exposure is due to the processing of the large data volume generated by the TpxCam sensor.

In total, we performed 130 optimization steps over a duration of 17 hours, representing a speed improvement by a factor of $3$ compared to our previous setup using a SPAD camera. As shown in Figure~\ref{TPX_speed_up}b, the target pixel intensity increases throughout the optimization process, demonstrating the expected enhancement. 

The scattering layer used in this experiment was generated by displaying a random phase pattern directly on the SLM. The resulting sum-coordinate projection is shown in Figure~\ref{TPX_speed_up}c. After optimization, the corrected phase pattern significantly narrows the correlation peak compared to the initial distribution, as illustrated in Figure~\ref{TPX_speed_up}d.

\begin{figure}[h!]
    \centering
    \includegraphics[width = 0.8\textwidth]{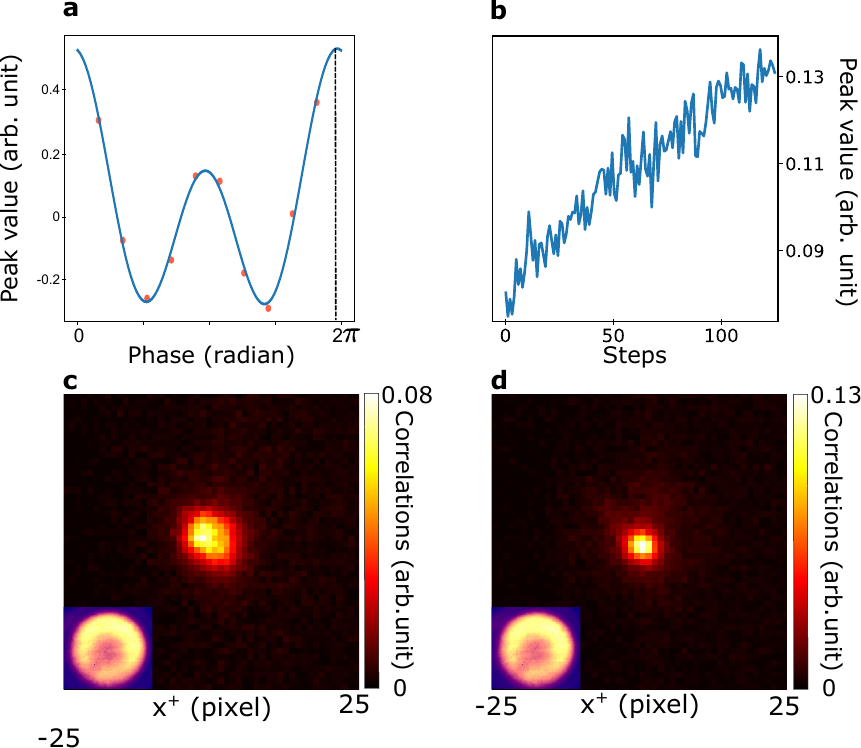}
\caption{
\textbf{Acceleration of non-classical optimization using the TpxCam sensor.}
\textbf{a,} Modulation of the target pixel value for $10$ phases steps between $0$ and $2\pi$.
\textbf{b,} Evolution of the coincidence count at the target pixel over 130 optimization steps. 
\textbf{c,} Measured sum-coordinate projection for a random phase mask used to emulate a scattering medium.
\textbf{d,} Measured sum-coordinate projection obtained after optimization, showing a  focus at the target location. Left bottom inserts are the two-photon intensity at the TpxCam plane.  
}
\label{TPX_speed_up}
\end{figure}

 \section{Non-classical optimization through thicker scattering media and different choice of the optimization target ($\Gamma_{kl}$ or $\Gamma^{+}_T$).}
 \label{section9}

To maintain a sufficient SNR during our measurements, the experimental results shown in Figures of the manuscript were obtained using thin scattering media. However, using a thin medium is not a prerequisite for implementing non-classical optimization. Furthermore, directly optimizing \( \Gamma_{kl} \) also works, provided the experimental SNR is sufficient. In this section, we present both simulation and experimental results using \( \Gamma_{kl} \) as an optimization target for both thin and thick scattering media, confirming that our conclusions hold regardless of the type of scattering medium. In particular, we demonstrate that the optimization also converges to phase masks that do not focus classical coherent light. 

\begin{itemize}
    \item Figure~\ref{FigSM6}a-d show simulation results for optimizing the coincidence rate between two arbitrarily selected output modes of a thin scattering medium. The input state incident on the SLM is simulated using a Double-Gaussian model~\cite{fedorov_gaussian_2009}, with parameters experimentally determined in Section VII. The transmission matrix used in the simulations was experimentally measured using classical coherent light and a thin scattering medium (one parafilm layer), using the method described in Ref.~\cite{popoff_measuring_2010}. Figure~\ref{FigSM6}a shows the enhancement of the coincidence rate between two output pixels, \(k\) and \(l\), with respective positions \((x_k, y_k) = (10, 15)\) and \((x_l, y_l) = (17, 42)\), as a function of the number of steps in the iterative algorithms. After optimization, \(\Gamma_{kl}\) reveals an enhancement of the coincidence rate around specific positions (Figure~\ref{FigSM6}c), confirming the success of the optimization. 
    %The focus point size is center to the target point \((x_l, y_l) \), the surrounding illuminated pixels are typically of the size of the speckle grain, observed in Figure ~\ref{FigSM6}b). 

    \item Figures~\ref{FigSM6}e-h and \ref{FigSM6}i-l show similar simulation results for a thicker scattering medium and a 'perfectly random' scattering medium, respectively. To simulate the thicker medium, we used an experimentally measured transmission matrix of a six-layer parafilm stack. For the 'perfectly random' scattering medium, the scattering matrix was generated using MATLAB's \texttt{rand} function. In both cases, the iterative optimization successfully enhanced the coincidence rate between the selected output pair of pixels (Figs.~\ref{FigSM6}g and k). The intensity images obtained after optimization, when the photon pair source was replaced by classical coherent light, do not show any focusing points for either situation (Figs.~\ref{FigSM6}h and j).

    \item Figures~\ref{FigSM6bis}a-c show hybrid experiment-simulation results for a thicker scattering medium. In this case, a transmission matrix of a six-layer parafilm stack was first measured using classical coherent light. Then, the optimization of \( \Gamma_{kl} \) was simulated on the computer using a two-photon state at the input, resulting in a phase mask shown in Figure~\ref{FigSM6bis}a and the correlation image in Figure~\ref{FigSM6bis}b. It confirms the success of the maximization of the coincidence rate between modes \(k\) and \(l\). This phase mask was then programmed on the SLM of the experiment operated with classical coherent light. This leads to a speckle pattern at the output, confirming again experimentally that this phase mask does not refocus the classical light.

    \item Figures~\ref{FigSM6bis}d-f show results similar to those obtained in Figures 4b-d of the manuscript, but with optimization on \( \Gamma_{kl} \) instead of the sum-coordinate projection. This experiment reproduces the correlations that would be obtained with a non-entangled state using classical coherent light, by multiplying the intensity values at the output i.e. \( \Gamma_{kl} = |E_k^{(o)}|^2 |E_l^{(o)}|^2 \). In this case, there is sufficient SNR to conduct the experiment without relying on simulations. Figure~\ref{FigSM6bis}d shows the optimal phase mask obtained. Figure~\ref{FigSM6bis}e shows the correlation image \( \Gamma_{kl} \), revealing the focii at the modes \(k\) and \(l\). Finally, Figure~\ref{FigSM6bis}f shows the intensity image where two focii also appear at modes \(k\) and \(l\), further confirming that the optimization performed on the correlations of a non-entangled state inevitably leads to focusing of classical light on the output modes in intensity.
\end{itemize}

 These results confirm that, even though for practical reasons we used our algorithm to optimize the correlations at an arbitrary position of the sum-coordinate projection of \(\Gamma\), our approach also works if we optimize the coincidences rate between a single pair of pixels, regardless of the complexity of the scattering media.

\begin{figure} [h!]
    \centering
    \includegraphics[width = 0.95\textwidth]{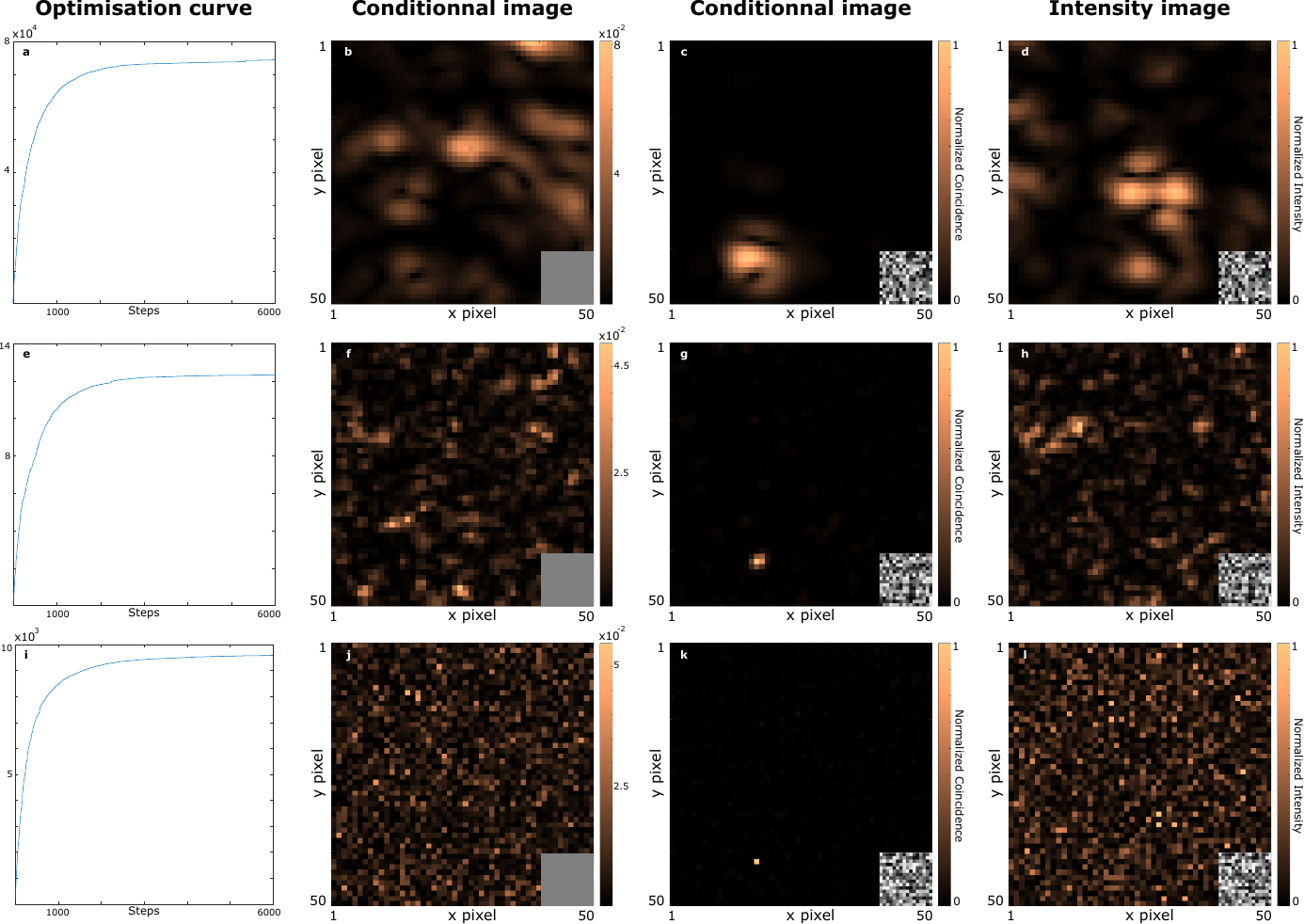}
    \caption{\justifying
\textbf{Simulation of focusing entangled photons into two arbitrary spatial modes for thin and thick scattering media.}
\textbf{a-d,} The simulation is performed using an experimentally measured scattering matrix of a thin scattering medium (one layer of parafilm) and an input state with the same properties as the one used in our experiment.  
\textbf{a,} Optimization curve using \(\Gamma_{kl}\) as a target, with respective positions \((x_k, y_k) = (10, 15)\) and \((x_l, y_l) = (17, 42)\).  
\textbf{b,} Conditional image \(\Gamma_{x,y|l}\) with a flat phase mask on the SLM.  
\textbf{c,} Conditional image \(\Gamma_{x,y|l}\) at the end of the optimization (optimal phase mask in inset).  
\textbf{d,} Output intensity observed after replacing photon pairs with classical coherent light, showing no focus.  
\textbf{e-h,} The simulation is performed using an experimentally measured scattering matrix of a thick scattering medium (six layers of parafilm) and an input state with the same properties as the one used in our experiment.  
\textbf{e,} Optimization curve using \(\Gamma_{kl}\) as a target, with respective positions \((x_k, y_k) = (10, 15)\) and \((x_l, y_l) = (17, 42)\).  
\textbf{f,} Conditional image \(\Gamma_{x,y|l}\) with a flat phase mask on the SLM.  
\textbf{g,} Conditional image \(\Gamma_{x,y|l}\) at the end of the optimization (optimal phase mask in inset).  
\textbf{h,} Output intensity observed after replacing photon pairs with classical coherent light, showing no focus.  
\textbf{i-l,} The simulation is performed using a scattering matrix generated using MATLAB's \texttt{rand} function, modeling a perfectly random scattering medium, and an input state with the same properties as the one used in our experiment.  
\textbf{i,} Optimization curve using \(\Gamma_{kl}\) as a target, with respective positions \((x_k, y_k) = (10, 15)\) and \((x_l, y_l) = (17, 42)\).  
\textbf{j,} Conditional image \(\Gamma_{x,y|l}\) with a flat phase mask on the SLM.  
\textbf{k,} Conditional image \(\Gamma_{x,y|l}\) at the end of the optimization (optimal phase mask in inset).  
\textbf{l,} Output intensity observed after replacing photon pairs with classical coherent light, showing no focus.  
All the transmission matrices used in these simulations connect \(16 \times 16\) input modes (SLM macropixels) to \(100 \times 100\) output modes (CCD camera pixels). To reduce computational memory requirements, we focus exclusively on the central \(50 \times 50\) pixels region at the output. }
    \label{FigSM6}
\end{figure}

\begin{figure} [h!]
    \centering
    \includegraphics[width = 0.95\textwidth]{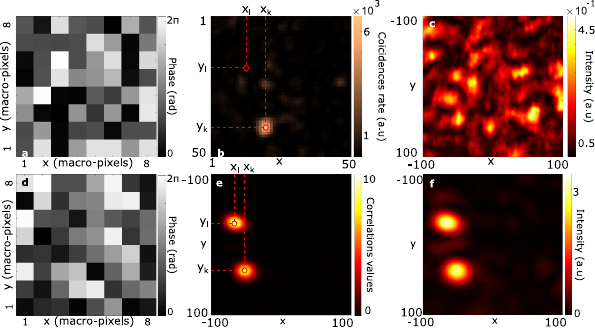}
    \caption{\justifying
    \textbf{Simulation and experiment of focusing entangled and non-entangled photons into two arbitrary spatial modes for a thick scattering medium.}
\textbf{a-c} The simulation is performed using an experimentally measured scattering matrix of a thick scattering medium (six-layer parafilm stack) and an input state with the same properties as the one used in our experiment.  \(\Gamma_{kl}\) is used as a target, with respective positions \((x_k, y_k) = (10, 15)\) and \((x_l, y_l) = (17, 42)\). 
\textbf{a,} Optimal phase mask obtained after optimization.
\textbf{b,} Conditional image \(\Gamma_{x,y|l}\) (simulation).
\textbf{c,} Output intensity observed after programming the phase mask of the SLM and performing the experiment with classical coherent light, showing no focus. 
\textbf{d-f} To reproduce the result that would be obtained using a non-entangled state, an experiment is performed using classical coherent light propagating through the scattering medium. The optimization is performed using the product of the intensities at positions \((x_k, y_k) = (10, 15)\) and \((x_l, y_l) = (17, 42)\) as the target. The experimental conditions are the same as those used to obtain the results shown in Figures 4b-d of the manuscript. 
\textbf{d,} Optimal phase mask obtained after optimization.
\textbf{e,} Conditional image \(\Gamma_{x,y|l}\) (experiment).
\textbf{f,} Output intensity, showing two focii (experiment).}
    \label{FigSM6bis}
\end{figure}

\section{Reproducibility of the results}

To verify that the main conclusion of our study — non-classical optimization leads to a phase mask that does not refocus classical light — is not an isolated result, we have of course (1) reproduced the optimization experiment with several scattering media (including some shown in Figure~\ref{FigSM9} in the next section), but also (2) conducted numerous simulations, as the latter are much faster to perform than the experiments. 

For example, we performed a batch of 10 additional simulations. In each simulation, the target was set to the central pixel of the sum-coordinate projection, and the input state matched the one used in the experiment. The scattering matrix used for these simulations corresponds to two layers of parafilm separated by \(2\)cm.
Figures~\ref{FigSM8}a-c show the intensity images obtained at the end of the optimization for three of these simulations, each showing a distinct speckle pattern. Figures~\ref{FigSM8}d-f show the respective sum-coordinate projections, all of which exhibit a strong peak at the central pixel, confirming the success of the optimization.

We observe, as expected, that when the non-classical optimization is repeated multiple times for a given medium and input state, a non-classical solution is always obtained. However, it is interesting to note that these solutions are always different from one another.
To quantitatively assess these differences, we calculate the similarity between all possible pairs of intensity and sum-coordinate images obtained using MATLAB's \texttt{Corr2} function. The average similarity value between each sum-coordinate projection obtained at the end of the optimization is \( C_q=0.971 \pm 0.007\), confirming a high degree of consistency among the images (i.e. all the sum-coordinate projection at the end of the optimization look the same). In contrast, for the classical intensity images, the average similarity is \( C_c=0.150 \pm 0.032\), confirming significant differences between the non-classical solution obtained (i.e. all the classical intensity images obtained at the end of the optimization are different). The different optimization simulations used to compute the correlation value are illustrated in Fig. \ref{FigSM9}. The convergence of the target pixel value from different initial conditions toward a common asymptotic limit suggests the presence of multiple local minima with close energy.  

\begin{figure} [h!]
    \centering
    \includegraphics[width = 1\textwidth]{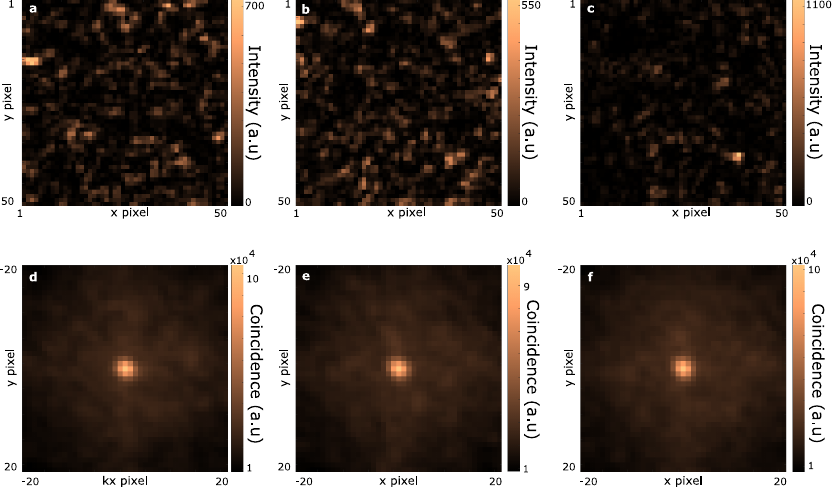}
    \caption{\justifying \textbf{Simulations to assess the consistency of the experiment.} \textbf{(a-c)} Intensity images and \textbf{(d-f)} obtained at the end of the optimization process for three simulations performed using the same scattering matrix and input state.} 
    \label{FigSM8}
\end{figure}

\begin{figure} [h!]
    \centering
    \includegraphics[width = 0.6\textwidth]{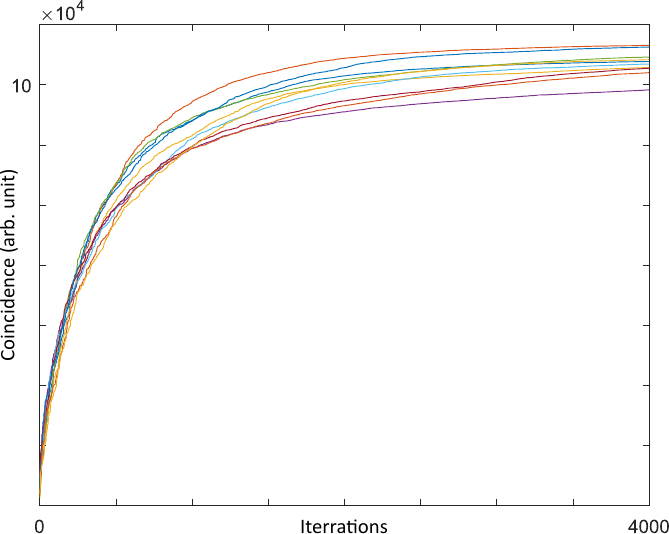}
    \caption{\justifying \textbf{Optimization curves for various initial phase masks.} Each curve represents an optimization process using the same quantum state properties and transmission matrix, but with different initial configurations of the SLM mask.}  
    \label{FigSM9}
\end{figure}

\section{Details on the non-entangled pure and mixed states}
\label{detailstates}

In the manuscript and in section~\ref{nonclaoptiwithsep}, we compare experimental results obtained with a two-photon entangled state to those obtained with two separable states. For a fair comparison, the separable states are chosen to reproduce, in the absence of the scattering medium, the same correlations in the camera plane as those of the entangled state. This section details the analytical form of the selected states and their properties.

\subsection{Pure separable state}

The first state chosen is a pure separable two-photon state. Its two-photon wavefunction in the SLM plane is written as $\psi(\vec{r_1},\vec{r_2}) = \phi(\vec{r_1}) \chi(\vec{r_2})$, where $\phi$ and $\chi$ are the optical fields associated with each photon. To determine $\phi$ and $\chi$, we impose the condition that the state must reproduce the same sum-coordinate projection $\Gamma^+$ as the entangled state when detected in the camera plane without the scattering medium i.e.
$\Gamma^+(\vec{r_+}) = A \exp{ \frac{-2 |\vec{r_+}|^2 } {{\lambda^2 f^2 \sigma_{\vec{k}}^2} } }$,
where $A$ is a normalization constant and $f$ is the effective focal length of the Fourier transform linking the input and output planes (see Equation~\eqref{eqGaussianSum} for details). 
To fullfill this condition, let's consider the following choice (Equation (24) in the manuscript): 
\begin{equation}
\label{choiceeq}
    \phi(\vec{r}) = \chi(\vec{r}) = A \exp{-\frac{|\vec{r}|^2 \sigma_{\vec{k}}^2}{8 }},
\end{equation} 
where $A$ is a normalization constant. To verify it, we propagate the corresponding two-photon wavefunction $\psi$ to the camera plane:
\begin{eqnarray}
\psi'(\vec{r'_1},\vec{r'_2})  &=& \iint \psi(\vec{r_1},\vec{r_2}) e^{i \frac{2 \pi}{\lambda f} \vec{r_1} \vec{r'_1}} e^{i \frac{2 \pi}{\lambda f} \vec{r_2} \vec{r'_2}} d \vec{r_1} d\vec{r_2} \\
&=& \int \exp{-\frac{|\vec{r_1}|^2 \sigma_{\vec{k}}^2}{8 }} e^{i \frac{2 \pi}{\lambda f} \vec{r_1} \vec{r'_1}} d\vec{r_1} \int \exp{-\frac{|\vec{r_2}|^2 \sigma_{\vec{k}}^2}{8 }}  e^{i \frac{2 \pi}{\lambda f} \vec{r_2} \vec{r'_2}}  d\vec{r_2} \\
&=& A \exp{-\frac{2|\vec{r'_1}|^2 }{\sigma_{\vec{k}}^2 \lambda^2 f^2}} \exp{-\frac{2|\vec{r'_2}|^2  }{\sigma_{\vec{k}}^2 \lambda^2 f^2}}.  
\end{eqnarray}
Then, we calculate the corresponding sum-coordinate projection:
\begin{eqnarray}
\Gamma^+(\vec{r_+})  &=& \int \Gamma(\vec{r_+}-\vec{r},\vec{r}) d \vec{r} \\
&=& \int |\psi'(\vec{r_+}-\vec{r},\vec{r})|^2 d \vec{r} \\
&=& A \int \exp{-\frac{4 |\vec{r_+}-\vec{r}|^2  }{\sigma_{\vec{k}}^2 \lambda^2 f^2}} \exp{-\frac{4 |\vec{\vec{r}}|^2  }{\sigma_{\vec{k}}^2 \lambda^2 f^2}} d \vec{r} \\
&=& A \exp{-\frac{2 |\vec{r_+}|^2}{\sigma_{\vec{k}}^2 \lambda^2 f^2} } \int \exp{-\frac{4}{\sigma_{\vec{k}}^2 \lambda^2 f^2} |\vec{r_+}/\sqrt{2}- \sqrt{2} \vec{r}|^2} d \vec{r} \\
&=& B \exp{-\frac{ 2|\vec{r_+}|^2}{\sigma_{\vec{k}}^2 \lambda^2 f^2} },
\end{eqnarray}
where $B$ is a normalization constant. The above calculation demonstrates that the choices made in Equation (24) of the manuscript and in Equation~\eqref{choiceeq} allow us to reproduce the same sum-coordinate projection as that obtained with the entangled state in the absence of the medium.

\subsection{Mixed separable state}

The second state chosen is a mixed separable state. In this case, we impose the same condition: the state must produces the same sum-coordinate projection $\Gamma^+$ than the entangled state when it is detected in the camera plane without medium i.e. $\Gamma^+(\vec{r_+}) = A \exp{ \frac{-2 |\vec{r_+}|^2 } {{\lambda^2 f^2 \sigma_{\vec{k}}^2} } }$, where $A$ is a normalization constant and $f$ is the effective focal length associated with the Fourier transform linking the input plane to the output plane (see Equation~\eqref{eqGaussianSum} for more details). As detailed in the manuscript, a separable mixed state can be written under the general form:
\begin{equation}
    \rho = \sum_j p_j \ket{\psi^j} \bra{\psi^j},
\end{equation}
where $\ket{\psi^j}$ is a pure state described by a separable two-photon wavefunction $\psi^j = \phi^j \chi^j$, and $p_j$ is its probability. In our calculations, we use a continuous-variable from of the above definition:
\begin{equation}
    \rho = \int p_{\vec{q}} \ket{\psi^{\vec{q}}} \bra{\psi^{\vec{q}}} d\vec{q}.
\end{equation}
By definition, the second-order correlation function $\Gamma$ associated with this mixed state is:
\begin{equation}
    \Gamma(\vec{r_1},\vec{r_2}) = \int p_{\vec{q}} |\psi^{\vec{q}} (\vec{r_1},\vec{r_2})|^2 d \vec{q}.
\end{equation}

We first consider the mixed state in the output (camera) plane, denoted $\rho'$. To ensure that this state exhibits the same sum-coordinate projection $\Gamma^+$ as the entangled state, a sufficient (and strong) condition is that it shares the same second-order correlation function $\Gamma$. This condition is satisfied by making the following choices:
\begin{eqnarray}
    p_{\vec{q}} &=& A\\
    \phi^{\vec{q}}(\vec{r'_1}) &=&  \exp{- \left | \frac{\vec{r'_1}}{\lambda f}-\vec{q} \right|^2 \sigma_{\vec{r}}^2} \exp{\frac{-\left | \frac{\vec{r'_1}}{\lambda f}+\vec{q} \right|^2 }{\sigma_{\vec{k}^2}}} \\
    \chi^{\vec{q}}(\vec{r'_2}) &=& \delta \left( \frac{\vec{r'_2}}{\lambda f} - \vec{q} \right),
\end{eqnarray}
where $A$ is a constant. To verify it, we calculate the sum-coordinate projection:
\begin{eqnarray}
\Gamma^+(\vec{r_+})  &=& \int \Gamma(\vec{r_+}-\vec{r},\vec{r}) d \vec{r} \\
&=& \iint p_{\vec{q}} |\psi^{\vec{q}}(\vec{r_+}-\vec{r},\vec{r})|^2  d \vec{q} d \vec{r} \\
&=& A \iint \exp{- 2 \left | \frac{\vec{r_+-\vec{r}}}{\lambda f}-\vec{q} \right|^2 \sigma_{\vec{r}}^2} \exp{\frac{-2 \left | \frac{\vec{r_+-\vec{r}}}{\lambda f}+\vec{q} \right|^2 }{\sigma_{\vec{k}^2}}} \delta \left( \frac{\vec{r}}{\lambda f} - \vec{q} \right)  d \vec{q} d \vec{r}  \\
&=& A \int \exp{  \frac{-2|\vec{r_+}-2 \vec{r}|^2 \sigma_{\vec{r}}^2}{\lambda^2 f^2} } \exp{ \frac{-2|\vec{r_+}|^2 } {{\lambda^2 f^2 \sigma_{\vec{k}}^2} } }  d \vec{r}  \\
&=& B \exp{-\frac{ 2|\vec{r_+}|^2}{\sigma_{\vec{k}}^2 \lambda^2 f^2} }.
\end{eqnarray}
These calculations confirm that the construction of $\rho'$ yields the same sum-coordinate projection $\Gamma^+$ as the entangled state.

To obtain the expression of the mixed state in the input (SLM) plane, denoted $\rho$, we back-propagate the mixed state $\rho'$ from the output plane. This operation corresponds to applying an inverse Fourier transform to each two-photon wavefunction that composes $\rho'$:
\begin{eqnarray}
    \psi^{\vec{q}}(\vec{r_1},\vec{r_2}) &=& \iint \phi^{\vec{q}} (\vec{r'_1}) \chi^{\vec{q}} (\vec{r'_2}) e^{-i \frac{\vec{r_1} \vec{r'_1}}{\lambda f}} e^{-i \frac{\vec{r_1} \vec{r'_1}}{\lambda f}} d\vec{r'_1} d\vec{r'_2} \\
    &=& A \int \phi^{\vec{q}} (\vec{k_1}) e^{-i \vec{r_1} \vec{k_1}} d\vec{k_1} \int \chi^{\vec{q}} (\vec{k_2}) e^{-i \vec{r_2} \vec{k_2}}  d\vec{k_2} \\
    &=& A \int \exp{- \left | \vec{k_1}-\vec{q} \right|^2 \sigma_{\vec{r}}^2} \exp{\frac{-\left | \vec{k_1}+\vec{q} \right|^2 }{\sigma_{\vec{k}^2}}} e^{-i \vec{r_1} \vec{k_1}} d\vec{k_1} \int \delta \left( \vec{k_2} - \vec{q} \right) e^{-i \vec{r_2} \vec{k_2}}  d\vec{k_2} \\
    &=& B \exp{- \frac{|\vec{r_1}|^2 \sigma_{\vec{k}}^2}{4(1 + \sigma_{\vec{r}}^2 \sigma_{\vec{k}}^2)}} e^{-i \vec{q} \vec{r_1}} e^{-i \vec{q} \vec{r_2}}.
\end{eqnarray}
We thus find the following definition for $\rho$ (Equations (25), (26) and (27) of the manuscript): 
\begin{align}
 \label{eqMixedx1} \phi^{\vec{q}}(\vec{r_1}) &= \exp{- \frac{|\vec{r_1}|^2 \sigma_{\vec{k}}^2}{4(1 + \sigma_{\vec{r}}^2 \sigma_{\vec{k}}^2)}} e^{-i \vec{q} \vec{r_1}} \\
 \label{eqMixedx2} \chi^{\vec{q}}(\vec{r_2}) &= e^{-i \vec{q} \vec{r_2}} \\ 
 \label{eqMixedx3} p_{\vec{q}} &= D, 
\end{align}
where $D$ is a constant chosen to ensure the proper normalization of the state.

\section{Details on the energy minimization of a multi-spin Hamiltonian}
\label{Isinsection}

\subsection{Demonstration of Equation (28) in the manuscript}

To demonstrate Equation (28) in the manuscript, we proceed step by step:

 \subsubsection{Start from Equation (1)}

In our experimental configuration, we use a single SLM to shape the input two-photon field. As a result, the term \(e^{i \theta_{nm}}\) in Equation (1) of the manuscript is replaced by the product \(e^{i \theta_{n}} e^{i \theta_{m}}\), where \(\theta_{n}\) is the phase value set at the input spatial mode \(n\) (i.e. the \(n^{\text{th}}\) SLM macropixel). Then, considering a phase encoding in \(0\) and \(\pi/2\), Equation (1) simplifies as follows:
\begin{equation}
    \Gamma_{kl} = \left | \sum_{m,n} t_{km} t_{ln} \psi_{mn} \epsilon_m \epsilon_n \right |^2,
    \label{equ13}
\end{equation}
where $\epsilon_n \in \{1,i \}$. 

 \subsubsection{Approximation $\psi_{mn} \approx \delta_{m-n} + \alpha \delta_{m-n+1}$}

To proceed, we use the approximation $\psi_{mn} \approx \delta_{m-n} + \alpha, \delta_{m-n+1}$, which is valid under our experimental conditions, as demontrated in the following.

In our experiment, each input spatial mode corresponds to an SLM macropixel composed of \(37 \times 37\) pixels, which corresponds to a square of approximately \(300 \times 10^{-5} \, \text{m}\) width. On the other hand, the position correlation width in the SLM plane is \(\sigma_{\vec{r}} \approx 2.9 \cdot 10^{-5} \, \text{m}\). Since the position correlation width is much smaller than the SLM macropixel width, we can approximate that the discretized two-photon wave function \(\psi_{nm}\) in the macropixel basis mainly contains nonzero coefficients along its diagonal, \(\psi_{nn} \neq 0\), and along the off-diagonals associated with coupling between directly neighboring pixels, \(\psi_{nn+1} \neq 0\), while other terms are negligible. The intensity of the off-diagonal coefficients relative to the diagonal ones is estimated by the ratio of the position correlation width to the SLM macropixel size, \(29/300 \approx 0.1\). We can thus simplify the discretized two-photon function (not normalized) as follows:  
\begin{equation}
\label{equapprox}
    \psi_{nm} \approx \delta_{n-m} + \alpha \delta_{n-m+1},
\end{equation}  
where \(n\) and \(m+1\) denote directly neighboring macropixels on the SLM. 

Note that Equation~\eqref{equapprox} has a very simplified form, as it corresponds to an SLM that would be one-dimensional. To account for the two-dimensional nature, i.e. the fact that each macropixel is actually coupled to 8 neighbors, additional terms must be included:  
\begin{eqnarray}
    \psi_{nm} &\approx& \delta_{n-m} + \alpha (\delta_{n-(m+1)}+\delta_{n-(m-1)}+\delta_{n-(m+\sqrt{N})}+\delta_{n-(m-\sqrt{N})}) \nonumber \\
    &+& \alpha' (\delta_{n-(m+\sqrt{N}-1)}+\delta_{n-(m+\sqrt{N}+1)}+\delta_{n-(m-\sqrt{N}-1)}+\delta_{n-(m-\sqrt{N}+1)}),
\end{eqnarray} 
where \(\alpha'\) is another coupling coefficient between diagonally neighboring pixels, and \(N\) is the number of SLM macropixels considered, forming a square of \(\sqrt{N} \times \sqrt{N}\). 
However, in the context of our work, there is no particular interest in using such a complex expression. To demonstrate Equation (28), we therefore simply consider Equation~\eqref{equapprox}.

 \subsubsection{Simplification of Equation~\eqref{equ13}}

Using the approximation $\psi_{mn} \approx \delta_{m-n} + \alpha \delta_{m-n+1}$, where $\alpha \approx 0.1$, Equation~\eqref{equ13} simplifies as follows:
\begin{equation}
    \Gamma_{kl} =\abs{\sum_n t_{kn} t_{ln} \epsilon_n^2 + \alpha \sum_n t_{kn} t_{l,n+1} \epsilon_n \epsilon_{n+1}}^2.
\end{equation}
Using the relation $\epsilon_n^2 = \sigma_n$, it simplifies in,
\begin{equation}
    \Gamma_{kl} = \abs{\sum_n t_{kn} t_{ln} \sigma_n + \alpha \sum_n t_{kn} t_{l,n+1} \epsilon_n \epsilon_{n+1}}^2.
\end{equation}
We then expand the absolute value of the sum of these terms into,
\begin{equation}
\begin{aligned}
\Gamma_{kl} = &\sum_{nm} t_{kn} t_{ln} t_{km}^* t_{lm}^* \sigma_n \sigma_m \\
&+ \alpha \Re\left(\sum_{nm} t_{kn} t_{ln} t_{km}^* t_{lm+1}^* \sigma_n \epsilon_m^* \epsilon_{m+1}^*\right) \\
&+ \alpha^2 \sum_{nm} t_{kn} t_{ln+1} t_{km}^* t_{lm+1}^* \epsilon_n \epsilon_{n+1} \epsilon_m^* \epsilon_{m+1}^*.
\label{equ14}
\end{aligned}
\end{equation}

\subsubsection{Simplification of the second term of Equation~\eqref{equ13}}

Expressing the $\epsilon_n$ in function of $\sigma_n$, one can demonstrate the following relation:
\begin{equation}
\epsilon_n\epsilon_m = \frac{1}{2} \left(\sigma_n + \sigma_m\right) + \frac{i}{2} \left(1 - \sigma_n \sigma_m\right).
\end{equation}
Inserting this into the second term of Equation~\eqref{equ14}, we obtain:
\begin{equation}
\Re\left(\sum_{nm} t_{kn} t_{ln} t_{km}^* t_{lm+1}^* \sigma_n \epsilon_m^* \epsilon_{m+1}^*\right) = \Re\left(\sum_{nm} t_{kn} t_{ln} t_{km}^* t_{lm+1}^* \left((\sigma_n\sigma_m + \sigma_n\sigma_{m+1} -i \left(\sigma_n - \sigma_n\sigma_m\sigma_{m+1}\right)\right)\right).
\end{equation}
Let $c_{nm}^{kl} = \Re\left(t_{kn} t_{ln} t_{km}^* t_{lm+1}^*\right)$ and $d_{nm}^{kl} = \Im\left(t_{kn} t_{ln} t_{km}^* t_{lm+1}^*\right)$, we can rewrite the second term as,
\begin{equation}
\Re\left(\sum_{nm} t_{kn} t_{ln} t_{km}^* t_{lm+1}^* \sigma_n \epsilon_m^* \epsilon_{m+1}^*\right) = \sum_{nm} c_{nm}^{kl} \left(\sigma_n\sigma_m + \sigma_n\sigma_{m+1}\right) + d_{nm}^{kl} \left(\sigma_n - \sigma_n\sigma_m\sigma_{m+1}\right).
\end{equation}

\subsubsection{Simplification of the third term of Equation~\eqref{equ13}}

Rewriting the third term of Equation~\ref{equ14} in terms of spins, we obtain:
\begin{equation}
\begin{aligned}
\sum_{nm} t_{kn} t_{ln+1} t_{km}^* t_{lm+1}^* \epsilon_n \epsilon_{n+1} \epsilon_m^* \epsilon_{m+1}^* = \sum_{nm} a_{nm}^{kl}\Re\left(\epsilon_n\epsilon_{n+1}\epsilon_m^*\epsilon_{m+1}^*\right) + b_{nm}^{kl}\Im\left(\epsilon_n\epsilon_{n+1}\epsilon_m^*\epsilon_{m+1}^*\right),
\end{aligned}
\end{equation}
where $a_{nm}^{kl} = \Re\left(t_{kn} t_{ln+1} t_{km}^* t_{lm+1}^*\right)$ and $b_{nm}^{kl} = \Im\left(t_{kn} t_{ln+1} t_{km}^* t_{lm+1}^*\right)$.
Then, we expand these terms and rewrite them in function of $\sigma_n$:
\begin{equation}
\begin{aligned}
\Re\left(4\epsilon_n\epsilon_{n+1}\epsilon_m\epsilon_{m+1}\right) &= 1 + \sigma_n\sigma_m + \sigma_n\sigma_{m+1} + \sigma_{n+1}\sigma_m + \sigma_{n+1}\sigma_{m+1} - \sigma_n\sigma_{n+1} - \sigma_m\sigma_{m+1} - \sigma_n\sigma_{n+1}\sigma_m\sigma_{m+1}\\
\Im\left(4\epsilon_n\epsilon_{n+1}\epsilon_m\epsilon_{m+1}\right) &= \sigma_m + \sigma_{m+1} - \sigma_n - \sigma_{n+1} + \sigma_n\sigma_{m}\sigma_{m+1} - \sigma_{n+1}\sigma_{m}\sigma_{m+1} - \sigma_n\sigma_{n+1}\sigma_m + \sigma_n\sigma_{n+1}\sigma_{m+1}.
\end{aligned}
\end{equation}

\subsubsection{General expression of $\Gamma_{kl}$}

We see that the coincidences between mode $k$ and $l$ written with terms of constants, one spin, two spins, three spins and four spins:
\begin{equation}
\Gamma_{kl} = C^{kl} + T_\sigma^{kl} + T_{\sigma^2}^{kl} + T_{\sigma^3}^{kl} + T_{\sigma^4}^{kl},
\label{equ15}
\end{equation}
where,
\begin{equation}
C^{kl} = \frac{\alpha^2}{4} \sum_{nm} a_{nm}^{kl},
\end{equation}
\begin{equation}
\begin{aligned}
T_\sigma^{kl} &= \sum_{nm} \alpha d_{nm}^{kl} \sigma_n + \frac{\alpha^2}{4} b_{nm}^{kl} \left(\sigma_m + \sigma_{m+1} - \sigma_n - \sigma_{n+1}\right)\\
&= \sum_{nm} \alpha d_{nm}^{kl} \sigma_n,
\end{aligned}
\end{equation}
where the last term cancels out due to the symmetry of $b$, $b^{kl}_{nm} = b^{kl}_{mn}$.
\begin{equation}
\begin{aligned}
T_{\sigma^2}^{kl} &= \sum_{nm} \alpha c_{nm}^{kl} \left(\sigma_n\sigma_m + \sigma_n\sigma_{m+1}\right) + \frac{\alpha^2}{4} a_{nm}^{kl} \left(\sigma_n\sigma_m + \sigma_n\sigma_{m+1} + \sigma_{n+1}\sigma_m + \sigma_{n+1}\sigma_{m+1} - \sigma_n\sigma_{n+1} - \sigma_m\sigma_{m+1}\right)\\
&= \sum_{nm} \alpha c_{nm}^{kl} \left(\sigma_n\sigma_m + \sigma_n\sigma_{m+1}\right) + \frac{\alpha^2}{4} a_{nm}^{kl} \left(\sigma_n\sigma_m + 2\sigma_n\sigma_{m+1} + \sigma_{n+1}\sigma_{m+1} - 2\sigma_n\sigma_{n+1}\right),
\end{aligned}
\end{equation}
where terms can be grouped because of the symmetry of $a$, $a^{kl}_{nm} = a^{kl}_{mn}$. This term can be written as:
\begin{equation}
T_{\sigma^2}^{kl} = \sum_{nm} J_{nm}^{kl} \sigma_n\sigma_m,
\end{equation}
where $J_{nm}^{kl} = \alpha c_{nm}^{kl} + \alpha c_{nm-1}^{kl} + \frac{\alpha^2}{4} a_{nm}^{kl} \left(1 - 2\delta_{m+1-n}\right) + \frac{\alpha^2}{2} a_{nm-1}^{kl} + \frac{\alpha^2}{4} a^{kl}_{n-1m-1}$. Then, the next term writes:
\begin{equation}
\begin{aligned}
T_{\sigma^3}^{kl} &= -\sum_{nm} \alpha d_{nm}^{kl} \sigma_n\sigma_m\sigma_{m+1} - \frac{\alpha^2}{4} b_{nm}^{kl} \left(\sigma_{n+1}\sigma_m\sigma_{m+1} - \sigma_n\sigma_{n+1}\sigma_m - \sigma_n\sigma_{n+1}\sigma_{m+1}\right)\\
&= -\sum_{nm} \alpha d_{nm}^{kl} \sigma_n\sigma_m\sigma_{m+1} - \frac{\alpha^2}{4} b_{nm}^{kl} \sigma_n\sigma_{n+1}\sigma_{m+1}\\
&= -\sum_{nmp}\Lambda_{nmp}^{kl}\sigma_n\sigma_m\sigma_p,
\end{aligned}
\end{equation}
where we have used the symmetry of $b$, $b^{kl}_{nm} = b^{kl}_{mn}$, and have defined $\Lambda_{nmp}^{kl} = \left(\alpha d_{nm}^{kl} + \frac{\alpha^2}{4} b_{nm}^{kl}\right)\delta_{m+1-p}$.
And finally, the last term writes:
\begin{equation}
\begin{aligned}
T_{\sigma^4}^{kl} &= \frac{\alpha^2}{4} \sum_{mn} a_{nm}^{kl} \sigma_n \sigma_m \sigma_{n+1} \sigma_{m+1}\\
&= \sum_{nmpq} Q_{nmpq}^{kl} \sigma_n \sigma_m \sigma_p \sigma_q,
\end{aligned}
\end{equation}
where we have defined $Q_{nmpq}^{kl} = \frac{\alpha^2}{4} a_{nm}^{kl} \delta_{n+1-p}\delta_{m+1-q}$. It should be noted that the terms $T_{\sigma^2}^{kl}$, $T_{\sigma^3}^{kl}$ and $T_{\sigma^4}^{kl}$ contain some constant terms - with respect to the spins — when $n=m$ or $n=m+1$. These terms can be absorbed into the constant term $C^{kl}$.

\subsubsection{From $\Gamma_{kl}$ to $\Gamma^{+}_T$}

In our work we do not directly optimize $\Gamma_{kl}$, but $\Gamma^{+}_T = \sum_{k=1}^M \Gamma_{T+k,T-k}$, which corresponds to an arbitrary pixel $T$ of the sum-coordinate projection of $\Gamma$. Using Equation~\eqref{equ15}, $\Gamma^{(+)}_T$ is then expressed as:
\begin{equation}
    \Gamma^{+}_T = C^T - \frac{1}{2} \sum_n K_n^T - \frac{1}{2} \sum_n J_{nm}^T \sigma_n \sigma_m - \frac{1}{2} \sum_n \Lambda_{nml}^T \sigma_n \sigma_m \sigma_l - \frac{1}{2} \sum_n Q_{nmlp}^T \sigma_n \sigma_m \sigma_l \sigma_p,
\end{equation}
where $C^T = \sum_{k=1}^M C^{T+k,T-k}$, $K_{n}^T = \sum_{k=1}^M K_n^{T+k,T-k}$, $J_{nm}^T = \sum_{k=1}^M J_{nm}^{T+k,T-k}$, $\Lambda_{nml}^T = \sum_{k=1}^M \Lambda_{nml}^{T+k,T-k}$ and $Q_{nmlp}^T = \sum_{k=1}^M Q_{nmlp}^{T+k,T-k}$. This result is Equation (28) of the manuscript. 

\subsubsection{Additional remarks}

In addition, it is interesting to note that if $\alpha = 0$, which corresponds to the case where the size of a SLM macropixel is much larger than the position correlation width $\sigma_{\vec{r}}$ of the photon pairs in the SLM plan, then Equation (28) of the manuscript simplifies into: 
  \begin{equation}
      \Gamma^+_T = - \frac{1}{2} \sum_n \left(\sum_{kl} Re(t_{(T+k)n} t_{(T-k)n} \bar{t}_{(T+k))m} \bar{t}_{(T-k)m}) \right) \sigma_n \sigma_m,
  \end{equation}
That is the Hamiltonian of a conventional spin-glass model with only spin-spin interaction, similar to those obtained with classical optical Ising machines~\cite{pierangeli_scalable_2021}.

Finally, as mentioned above, it is also important to note that the approximation \(\psi_{mn} \approx \delta_{m-n} + \alpha \delta_{m-n+1}\) assumes a one-dimensional SLM i.e. the coefficient \(\alpha\) only creates coupling between one direct neighboring SLM macropixel. However, in reality, since the SLM is bidimensional, each SLM macropixel couples with $8$ neighboring pixels, meaning that additional terms should be included. As a result, the coefficients \(C^T\), \(K^T_{n}\), \(J_{nm}^T\), \(\Lambda_{nml}^T\), and \(Q_{nmlp}^T\) would be modified accordingly, but the model would still retain a structure with multi-spin interactions \(\sigma_n \sigma_m \sigma_l\) and \(\sigma_n \sigma_m \sigma_l \sigma_p\), which is the key aspect of our demonstration.

{\subsection{Experimental simulation of an Ising model with multi-spin interactions}

Optimizing the coincidence rate at a target output position, for SLM pixels values restricted to be $0$ or $\pi/2$, corresponds to minimizing the energy of the Hamiltonian \(H\) described in the above section. Using our experimental system, we demonstrate such a process using \(8 \times 8 = 64\) spins (i.e. SLM input modes). Figure~\ref{fig 5}a shows the energy minimization curve, where the energy is simply defined as the negative of the target coincidence rate $\Gamma_T^+$ in the output sum-coordinate projection. To reduce it, each optimization step randomly selects $25\%$ of the \(8 \times 8\) spatial input modes (SLM macropixels) and limits the phase encoding to $0$ or \(\pi/2\), effectively flipping the spins i.e all values at $0$ are set to \(\pi/2\) and vice versa. After flipping, the energy is evaluated, and the new configuration is kept only if the energy decreases.  In our experiment, flipping $25\%$ of the spins ensures a sufficient signal-to-noise ratio for each measurement, though in theory, spins could be flipped individually. This fraction also determines the accuracy of the ground state found with the method, and explains the plateaus observed in the minimization curve in Figure~\ref{fig 5}a.  After the non-classical optimization, the sum-coordinate projection reveals a focused spot at the target position (Figure~\ref{fig 5}b), and the SLM phase pattern converges to a configuration corresponding to a local minimum of the energy landscape (Figure~\ref{fig 5}c). In this experiment, the scattering medium was created by programming a fixed random phase pattern on the SLM, with similar complexity than the layer of parafilm used in the other experiments. }

\begin{figure} [ht!]
    \centering
    \includegraphics[width = 1\columnwidth]{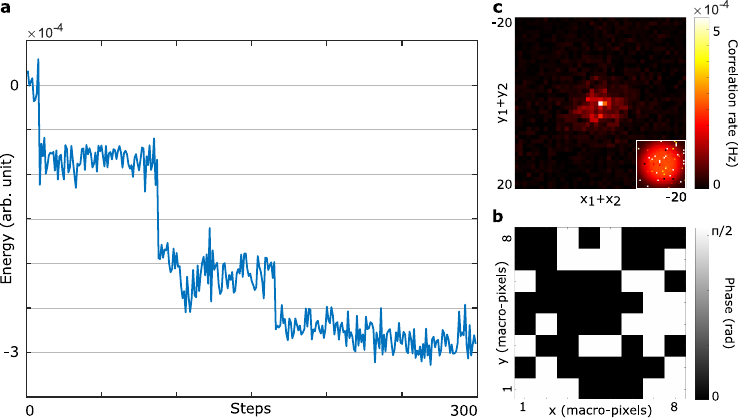}
\caption{\textbf{Optical simulation of an Ising model with multi-spin interactions.} \textbf{a,} Energy of the Ising model as the number of optimization steps. The optimization is performed with $8 \times 8$ SLM macropixels at the input, each taking only two values, $0$ or \(\pi/2\). \textbf{b,} Sum-coordinate projection $\Gamma^+$ measured at the end of the optimization, in which the central target pixel rises above the background. \textbf{c,} Optimal SLM phase mask reached after 300 steps of optimization (100 hours), corresponding to the ground state of the simulated system. The spins up (+1) and down (-1) correspond to the values of 0 and \(\pi/2\), respectively.}
    \label{fig 5}
\end{figure}

\section{Non-classical optimization with non-entangled states}
\label{nonclaoptiwithsep}

As detailed in the manuscript, `non-classical’ solutions - phase masks that refocus entangled photons without refocusing classical light - arise from the presence of entanglement in the input state. To support this claim, we compare our optimization results with those obtained using non-entangled states. For a fair comparison, these separable states are chosen to reproduce, in the absence of the scattering medium, spatial correlations in the camera plane similar to those observed experimentally with the entangled state. The analytical form of the two separable states considered here is described in section~\ref{detailstates} and in the Methods section of the manuscript.

\subsection{Experiment with a pure separable state}

{First, We consider the case of the pure two-photon state, which the factorizable wavefunction i.e. \( \psi_{nm} = \phi_n \chi_m \), where $\phi$ and $\chi$ are optical fields associated with each photon, is described by Equation (24) of the manuscript and Equation~\eqref{choiceeq}. With a pure separable state, Equation (3) of the manuscript simplifies as: 
\begin{equation}
    \Gamma_{kl} = \left | \sum_{n} t_{kn} \phi_{n} e^{i \theta_{n}} \right |^2 \left |  \sum_{m} t_{lm} \chi_{m} e^{i \theta_{m}} \right |^2.
\end{equation}}
In this case, maximizing the correlations between output modes $k$ and $l$ is equivalent to maximizing the product of the output intensities measured at modes $k$ and $l$ when two coherent classical fields, $\phi$ and $\chi$, are propagated through the medium. 

Experimentally, this situation is implemented by illuminating the SLM with a collimated classical coherent beam having a width similar to that of the SPDC light and by optimizing the product of the output intensities. Without the scattering medium, we thus observe a strong peak in the sum-coordinate projection (Fig.~\ref{fig 4}a). After propagation through the scattering medium, a diffuse halo appears (Fig.~\ref{fig 4}b). After {non-classical} optimization, the correlation peak is {retrieved} (Fig.~\ref{fig 4}c). Unlike the entangled case, however, a focus is also observed at the center of the output intensity image in Figure~\ref{fig 4}d. In this case therefore, the classical and non-classical optimization methods are equivalent. Note that this experiment was repeated several times to ensure the reproducibility of the result.

\begin{figure*}[!ht]
    \centering
    \includegraphics[width = 1\textwidth]{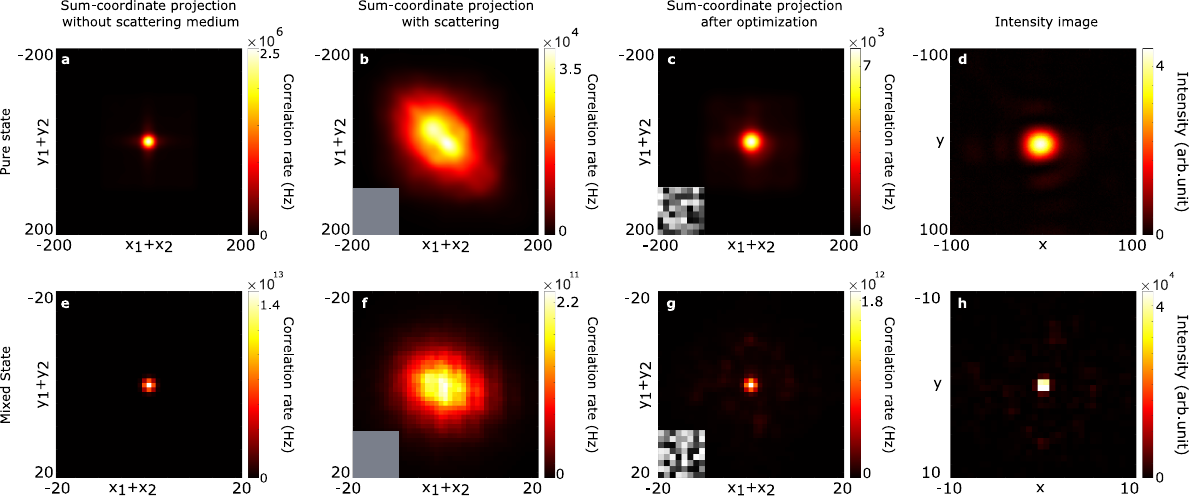}
\caption{{\textbf{Optimization results with non-entangled states.}  
Sum-coordinate projections show strong correlation peaks in the absence of scattering for both the pure \textbf{(a)} and mixed \textbf{(e)} separable states. When a scattering medium is introduced, this peak disappears and is replaced by a diffuse halo in both cases \textbf{(b,f)}. After applying the non-classical optimization procedure, the correlation peak is restored in the sum-coordinate projection \textbf{(c,g)}. Simultaneously, a focus also appears in the output intensity profiles \textbf{(d)} and \textbf{(h)} measured under classical coherent illumination, demonstrating that the optimization also converges to the classical solution. Optimal SLM patterns are shown in the insets.
Images \textbf{(a–d)} are experimental measurements obtained using a coherent classical light beam to mimic the properties of the pure separable state. Images \textbf{(e–h)} are results of numerical simulations, obtained by propagating the mixed separable state through a system replicating the optical setup of Figure~2f of the manuscript. Further details on the separable states and the simulations are provided in the Methods section of the manuscript, sections~\ref{detailstates}, ~\ref{sectionII} and~\ref{section9} .}}
    \label{fig 4}
\end{figure*}

\subsection{simulations with a mixed separable state}

 Second, we consider the case of a non-entangled mixed state, defined by its density matrix $\rho =  \sum_{j} p_j \left | \psi^j \right \rangle  \left \langle \psi^j \right |$, where $ \ket{\psi^j}$ is a separable two-photon state of factorizable wavefunction $\psi_{nm}^j = \phi^j_n \chi^j_m$, and $p_j$ its probability. As detailed in Methods and section~\ref{detailstates}, we choose $\phi^j$ and $\chi^j$ so that the mixed state reproduces the same spatial correlations (i.e. the same $\Gamma$ and $\Gamma^+$) as the entangled state without scattering, again providing a reliable point of comparison. 
 
 To test the optimization, the state $\rho$ is numerically propagated through a simulated optical system replicating the experimental setup shown in Figure~2f of the manuscript. In this case, the scattering matrix was generated from a random phase mask followed by a Fourier transform, producing a disordered medium with statistical properties similar to those observed experimentally. In addition, the state $\rho$ was generated according to the model defined in Equations (25-27) of the manuscript (same as Equations~\eqref{eqMixedx1},~\eqref{eqMixedx2} and~\eqref{eqMixedx3}), using the experimentally measured values of $\sigma_{\vec{k}}$ and $\sigma_{\vec{x}}$ to closely match the characteristics of the entangled state and the optical setup. In the absence of scattering medium, Figure~\ref{fig 4}e shows a strong peak at the center of the simulated sum-coordinate projection. 
 When a medium is introduced, this peak vanishes and is replaced by a diffuse halo (Fig.~\ref{fig 4}f).
 After non-classical optimization, a refocusing of the correlations on the sum-coordinate is observed (Fig.~\ref{fig 4}g). 
 However, the same optimal SLM pattern also leads to the refocusing of classical coherent light (Fig.~\ref{fig 4}h), indicating that in this case, the optimization is equivalent to a classical one.
 These results confirm that optimizing spatial correlations of non-entangled (pure or mixed) two-photon states consistently leads to the classical solution, underscoring the crucial role of entanglement in our scheme. Note that the simulations were repeated multiple times to ensure the reproducibility of the result.

 \end{document}